\documentclass[times]{fldauth}

\usepackage{moreverb}

\usepackage[colorlinks,bookmarksopen,bookmarksnumbered,citecolor=red,urlcolor=red]{hyperref}
\usepackage{color}
\usepackage{amssymb}
\usepackage{graphicx}
\usepackage{amsmath}
\usepackage{subfig}
\usepackage{booktabs}
\usepackage{tikz}
\usepackage{pgfplots}
\usepackage{afterpage}
\usepgfplotslibrary{external}
\tikzset{external/system call={lualatex \tikzexternalcheckshellescape -halt-on-error -interaction=batchmode -jobname "\image" "\texsource"}}
\usepackage{placeins}
\usepackage{morefloats}

\def\Xint#1{\mathchoice
{\XXint\displaystyle\textstyle{#1}}
{\XXint\textstyle\scriptstyle{#1}}
{\XXint\scriptstyle\scriptscriptstyle{#1}}
{\XXint\scriptscriptstyle\scriptscriptstyle{#1}}
\!\int}
\def\XXint#1#2#3{{\setbox0=\hbox{$#1{#2#3}{\int}$ }
\vcenter{\hbox{$#2#3$ }}\kern-.6\wd0}}

\def\dashint{\Xint-}

\newcommand\BibTeX{{\rmfamily B\kern-.05em \textsc{i\kern-.025em b}\kern-.08em
		T\kern-.1667em\lower.7ex\hbox{E}\kern-.125emX}}

\def\volumeyear{2017}

\begin{document}

\runningheads{R. Pethiyagoda, T. J. Moroney and S. W. McCue}{Efficient computation of two-dimensional steady free-surface flows}

\title{Efficient computation of two-dimensional \\ steady free-surface flows}

\author{Ravindra Pethiyagoda, Timothy J. Moroney and Scott W. McCue\footnote{School of Mathematical Sciences, Queensland University of Technology, Brisbane QLD 4001, Australia}\vphantom{d}$^{\text{,}}$\footnote{Email: scott.mccue@qut.edu.au}}

\address{School of Mathematical Sciences, Queensland University of Technology, Brisbane QLD 4001, Australia}

\begin{abstract}
We consider a family of steady free-surface flow problems in two dimensions, concentrating on the effect of nonlinearity on the train of gravity waves that appear downstream of a disturbance.  By exploiting standard complex variable techniques, these problems are formulated in terms of a coupled system of Bernoulli's equation and an integral equation.  When applying a numerical collocation scheme, the Jacobian for the system is dense, as the integral equation forces each of the algebraic equations to depend on each of the unknowns. We present here a strategy for overcoming this challenge, which leads to a numerical scheme that is much more efficient than what is normally employed for these types of problems, allowing for many more grid points over the free surface. In particular, we provide a simple recipe for constructing a sparse approximation to the Jacobian that is used as a preconditioner in a Jacobian-free Newton-Krylov method for solving the nonlinear system.
We use this approach to compute numerical results for a variety of prototype problems including flows past pressure distributions, a surface-piercing object and bottom topographies.
\end{abstract}


\maketitle

\vspace{-6pt}
\section{Introduction}
Steady nonlinear two-dimensional free-surface flow problems have been popular since the early 1980s, with computers allowing researchers to numerically solve the fully nonlinear problems using complex variable methods.  One well-cited example is Forbes and Schwartz~\cite{forbes82}, who considered the two-dimensional problem of flow of an ideal fluid past a semi-circular bottom obstruction. Numerical solutions show how a train of gravity waves develop on the free surface downstream from the disturbance, and how the waves themselves steepen as the size of the semi-circular disturbance increases.  The crests of the waves become sharper, the troughs broader, and the wavelength decreases when compared to that predicted by linear theory.  Since that time, this general approach has been adapted and extended in many studies for other bottom topographies with curved and straight boundaries (see, for example, \cite{binder13,binder06,binder08,dias02,forbes82a,king87,king90,vandenbroeck87,zhang96}).

Other well-studied two-dimensional examples include flow past a pressure distribution that is applied to the surface of the stream \cite{schwartz81}, flows past surface-piercing solid bodies \cite{farrow95,vandenbroeck80} and flows past submerged obstacles or singularities \cite{forbes85a,king89}.  These configurations are interesting in their own right, but also provide simple models for flow due to a steadily moving two-dimensional ship or submarine.  In each of these cases, a train of waves develops downstream from the disturbance.  The complex variable formulation and subsequent numerical solution provides insight into how the size and steepness of the waves depend on the nature of the disturbance.

The overall approach taken for all of these examples is essentially the same.  The Cauchy integral formula is used to enforce Laplace's equation, leaving an integral equation that holds on the free surface.  This equation is coupled with Bernoulli's equation, also applied to the free surface.  Discretising the equations and applying the integral equation at mesh points (or half-mesh points) leads to a system of algebraic equations, which can be solved using Newton's method.  The nonlocal nature of the integral equation results in each of the unknowns at the mesh points appearing in all of the algebraic equations and, as such, the Jacobian of the system is either fully dense or has a fully dense block.  For this reason, the number of mesh points used in practice is limited.

By considering a number of prototype problems, we illustrate shared properties of the Jacobians for each of these two-dimensional steady free-surface flows, such as the general block structure and the entries that depend on the integro-differential equation. Our core contribution is to demonstrate how the nonlinear systems of algebraic equations can be solved using a Jacobian-free Newton-Krylov method, which for a range of problems does not require the formation and factorisation of the dense Jacobian, and is much more efficient than standard schemes.  This approach is based on ideas proposed in our previous work for three-dimensional free-surface flows \cite{pethiyagoda14a}.  An important component of the scheme is the formation of a sparse preconditioner that is a sufficiently good approximation to the full Jacobian. The scheme allows for solutions on meshes with a larger number of collocation points and works equally well for all the problems treated.

The structure of the paper is as follows. In section \ref{2Dsec:GovEq} we formulate the problem of steady flow past a pressure distribution using the boundary integral approach mentioned above. This problem illustrates all of the key features we are concerned with. Section \ref{2Dsec:NumSch} is devoted to outlining our new numerical scheme. In sections \ref{2Dsec:infinitedepth} and \ref{2Dsec:finiteDepth} we demonstrate our approach via examples. Finally, in section \ref{2Dsec:Discussion} we discuss the significance of our study.

\section{Governing equations}
\label{2Dsec:GovEq}
\begin{figure}
\centering
\subfloat[$z$-plane]{\includegraphics[width=.46\linewidth]{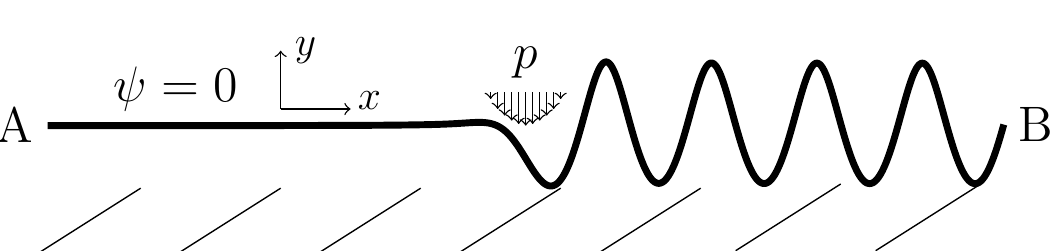}}\hspace{0.04\linewidth} 
\subfloat[$w$-plane]{\includegraphics[width=.46\linewidth]{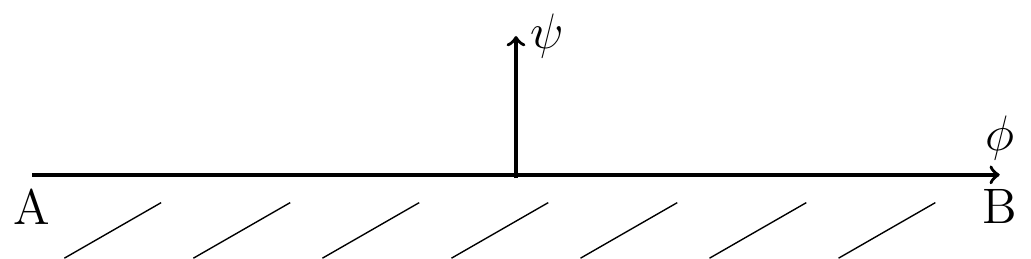}}
\caption{A schematic for flow past a pressure distribution, $p$, in an infinitely depth fluid. The physical plane is represented in part (a) while the flow field in part (b) is the lower half complex potential plane.}
\label{2Dfig:pressSchem}
\end{figure}

To begin, we consider a general two-dimensional steady free-surface flow problem for an ideal fluid of infinite depth with an unperturbed speed $U$ in the positive $x$-direction. The fluid is subject to a prescribed surface pressure distribution $p(\phi)$ applied to the surface of the fluid, where the velocity potential of the fluid $\phi(x,y)$ satisfies Laplace's equation in the fluid domain
\begin{equation}
\nabla^2\phi=0,\label{2Deq:laplace}
\end{equation}
and Bernoulli's equation on the surface $y=\eta(x)$
\begin{equation}
\frac{1}{2}|\nabla\phi|^2+\frac{y}{F_L^2}+p(\phi)=\frac{1}{2},\qquad\text{on }y=\eta(x),\label{2Deq:bern}
\end{equation}
where $F_L=U/\sqrt{gL}$ is the Froude number, $g$ is acceleration due to gravity and $L$ is a length scale related to the pressure distribution (more generally $L$ is some length scale relevant to the problem at hand). The pressure distribution has the effect of disturbing the free-surface, causing a train of gravity waves to develop downstream, as indicated in Figure \ref{2Dfig:pressSchem}(a). From a practical perspective, this pressure can be interpreted as mimicking a two-dimensional air cushion vehicle such as a hovercraft. This nonlinear free-surface problem, which can be considered a prototype two-dimensional flow, has received a variety of attention since the work of Schwartz~\cite{schwartz81}.

Considering the problem in the complex plane, we know that $w(z)=\phi+\mathrm{i}\psi$ is an analytic function of $z=x+\mathrm{i} y$, where $\psi$ is the streamfunction. We set $\psi=0$ on the free surface $y=\eta(x)$, thus mapping the fluid domain to the lower half-plane (Figure \ref{2Dfig:pressSchem}). Differentiating $w(z)$ with respect to $z$ gives
\begin{equation}
\frac{\mathrm{d}w}{\mathrm{d}z}=u-\mathrm{i} v=\mathrm{e}^{\tau-\mathrm{i}\theta},\label{2Deq:dwdz}
\end{equation}
where $u$ and $v$ are the horizontal and vertical fluid velocities, respectively. Here $\mathrm{e}^\tau$ is the fluid speed, while $\theta$ is the angle the streamline makes to the horizontal.

Because $\mathrm{d}w/\mathrm{d}z$ is an analytic function of $w$, we can use Cauchy's integral formula to perform a contour integral of $\log(\mathrm{d}w/\mathrm{d}z)$ over the lower half-plane. Taking the imaginary part leads to the Hilbert transform
\begin{equation}
\tau(\phi)=\frac{1}{\pi}\dashint_{-\infty}^{\infty}\frac{\theta(\phi')}{\phi'-\phi}\,\mathrm{d}\phi',\qquad -\infty<\phi<\infty.\label{2Deq:intEq}
\end{equation}
By enforcing (\ref{2Deq:intEq}), Laplace's equation (\ref{2Deq:laplace}) is identically satisfied. To satisfy Bernoulli's equation in terms of $\tau$ and $\theta$, we differentiate equation (\ref{2Deq:bern}) with respect to $\phi$, using the relation $|\nabla\phi|=\mathrm{e}^\tau$, to give
\begin{equation}
F_L^2\mathrm{e}^{3\tau}\frac{\mathrm{d}\tau}{\mathrm{d}\phi}+\sin\theta+F_L^2\mathrm{e}^{\tau}\frac{\mathrm{d}p}{\mathrm{d}\phi}=0,\qquad -\infty<\phi<\infty.\label{2Deq:bernComplex}
\end{equation}
Equations (\ref{2Deq:intEq}) and (\ref{2Deq:bernComplex}) are now our two governing equations for the two unknowns $\tau(\phi)$ and $\theta(\phi)$, that depend on the Froude number $F_L$ and the prescribed pressure distribution $p(\phi)$.

\section{Numerical scheme}
\label{2Dsec:NumSch}
\subsection{Collocation}
To tackle the problem outlined in section \ref{2Dsec:GovEq} computationally, we will construct a system of nonlinear equations to be solved with a Jacobian-free Newton-Krylov method. First, the domain from $-\infty<\phi<\infty$ is truncated to some finite domain. We then discretise the functions $\tau$ and $\theta$ along the truncated surface at the $N$ equally spaced mesh points $\phi_i$, for $i=1,\dots,N$, giving the values $\tau_i=\tau(\phi_i)$ and $\theta_i=\theta(\phi_i)$. The values $\tau_1$ and $\theta_1$ are given by the radiation condition far upstream. We are then left with $2(N-1)$ unknowns to be arranged in the vector \textbf{u},
\begin{equation}
\textbf{u}=[\tau_2,\dots,\tau_N,\theta_2,\dots,\theta_N]^T.\label{2Deq:unkowns}
\end{equation}
We require $2(N-1)$ equations to close this system. The first $N-1$ equations come from enforcing Bernoulli's equation (\ref{2Deq:bernComplex}) at the half-mesh points,
\begin{equation}
\tau_{i+1/2}=\frac{\tau_i+\tau_{i+1}}{2},\quad\theta_{i+1/2}=\frac{\theta_i+\theta_{i+1}}{2},\quad\frac{\mathrm{d}\tau_{i+1/2}}{\mathrm{d}\phi}=\frac{\tau_{i+1}-\tau_i}{\Delta\phi},\label{2Deq:halfmesh}
\end{equation}
for $i=1,\dots,N-1$ where $\Delta\phi$ is the mesh spacing. The final $N-1$ equations come from the integral equation (\ref{2Deq:intEq}), but first we deal with the singularity by the addition and subtraction of the term
\[
\frac{\theta(\phi)}{\pi}\dashint\frac{1}{\phi'-\phi}\,\mathrm{d}\phi',
\]
to give
\begin{equation}
\tau(\phi)=\frac{1}{\pi}\int_{\phi_1}^{\phi_N}\frac{\theta(\phi')-\theta(\phi)}{\phi'-\phi}\,\mathrm{d}\phi'+\frac{\theta(\phi)}{\pi}\ln\left|\frac{\phi_N-\phi}{\phi_1-\phi}\right|, \qquad -\infty<\phi<\infty.
\label{2Deq:intAltered}
\end{equation}
Now we can enforce equation (\ref{2Deq:intAltered}) at the half-mesh points (\ref{2Deq:halfmesh}) by moving all terms to the left hand side and computing the integral by the trapezoidal rule to form our final $N-1$ equations. We have then a system of $2(N-1)$ equations, $\textbf{F}(\textbf{u})=\textbf{0}$, for the $2(N-1)$ unknowns $\textbf{u}$ defined by (\ref{2Deq:unkowns}). We solve this system using a preconditioned Jacobian-free Newton-Krylov (JFNK) method.

\subsection{The Jacobian-free Newton-Krylov method}
The JFNK method is a Newton-like scheme where the preconditioned Generalised Minimum Residual (GMRES) algorithm is used for the typical Newton solve step \cite{brown90,saad86}. For clarity, the Newton solve step determines the search direction $\delta\textbf{u}_k$ by solving the system
\begin{equation}
\textbf{J}(\textbf{u}_k) \delta\textbf{u}_k = -\textbf{F}(\textbf{u}_k),
\label{2Deq:newtonstep}
\end{equation}
where $\textbf{J}(\textbf{u}_k)$ is the Jacobian evaluated at the $k\mathrm{th}$ iterate $\textbf{u}_k$. The GMRES algorithm solves equation (\ref{2Deq:newtonstep}) by first solving for $\textbf{z}$ in
\begin{align}
\textbf{J}(\textbf{u}_k)\textbf{P}^{-1}\textbf{z} &= -\textbf{F}(\textbf{u}_k), \label{2Deq:preGRMRES}
\end{align}
where
\begin{align}
\textbf{z}&=\textbf{P}\delta\textbf{u}_k,\label{2Deq:zsolve}
\end{align}
and $\textbf{P}$ is a preconditioner matrix. Equation (\ref{2Deq:preGRMRES}) is solved by projecting onto the preconditioned Krylov subspace of dimension $m$
\begin{align*}
\mathcal{K}_m(\textbf{J}_k\mathbf{P}^{-1},\textbf{F}_k)=\mathrm{span}\{\textbf{F}_k,\textbf{J}_k\mathbf{P}^{-1}\textbf{F}_k,\dots,(\textbf{J}_k\mathbf{P}^{-1})^{m-1}\textbf{F}_k\},
\end{align*}
where for simplicity we write $\textbf{J}_k=\textbf{J}(\textbf{u}_k)$, $\textbf{F}_k=\textbf{F}(\textbf{u}_k)$. The preconditioner matrix $\textbf{P}$, chosen such that $\textbf{P}\approx\textbf{J}_k$, is needed to keep the Krylov subspace dimension $m$ sufficiently small. Finally, we solve for $\delta\textbf{u}_k$ using (\ref{2Deq:zsolve}). The JFNK method is advantageous in that the full Jacobian does not need to be computed since its action on an arbitrary vector can be computed using a forward difference quotient~\cite{knoll04}. The preconditioner matrix, however, must be chosen to be easily computed and inverted to make effective use of the method.

\subsection{Preconditioning}\label{2Dsubsec:Precon}
By differentiating (\ref{2Deq:bernComplex}) and (\ref{2Deq:intAltered}) with respect to the unknowns $\tau_i$ and $\theta_i$ for $i=2,\dots,N$, we can easily compute the exact Jacobian $\textbf{J}_k$ by hand: details are provided in appendix~\ref{2Dsec:FormJac}. In the first instance, one may be inclined to use the full Jacobian at the initial Newton iterate as a preconditioner, updated only as often as required to keep the Krylov subspace dimension from growing too large. Unfortunately, this would require storing and factorising a $2(N-1)\times 2(N-1)$ matrix, which would restrict the maximum number of mesh points allowable on a given machine due to available system memory.

\begin{figure}
\centering
\subfloat[Full Jacobian]{\includegraphics[width=.46\linewidth]{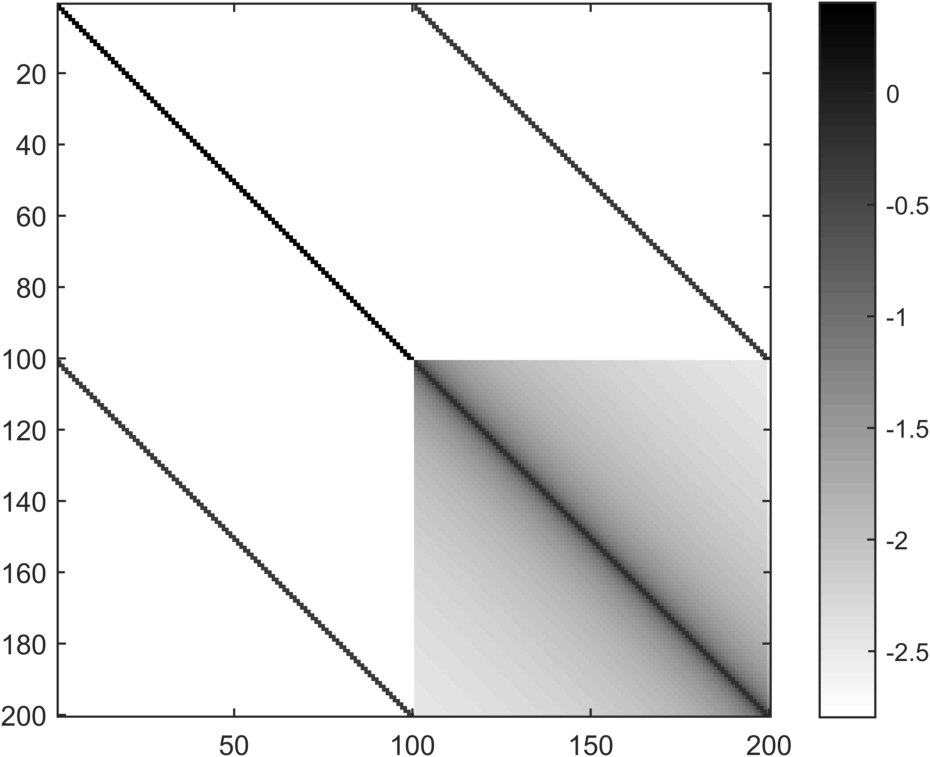}}\hspace{0.04\linewidth} 
\subfloat[Schur complement]{\includegraphics[width=.46\linewidth]{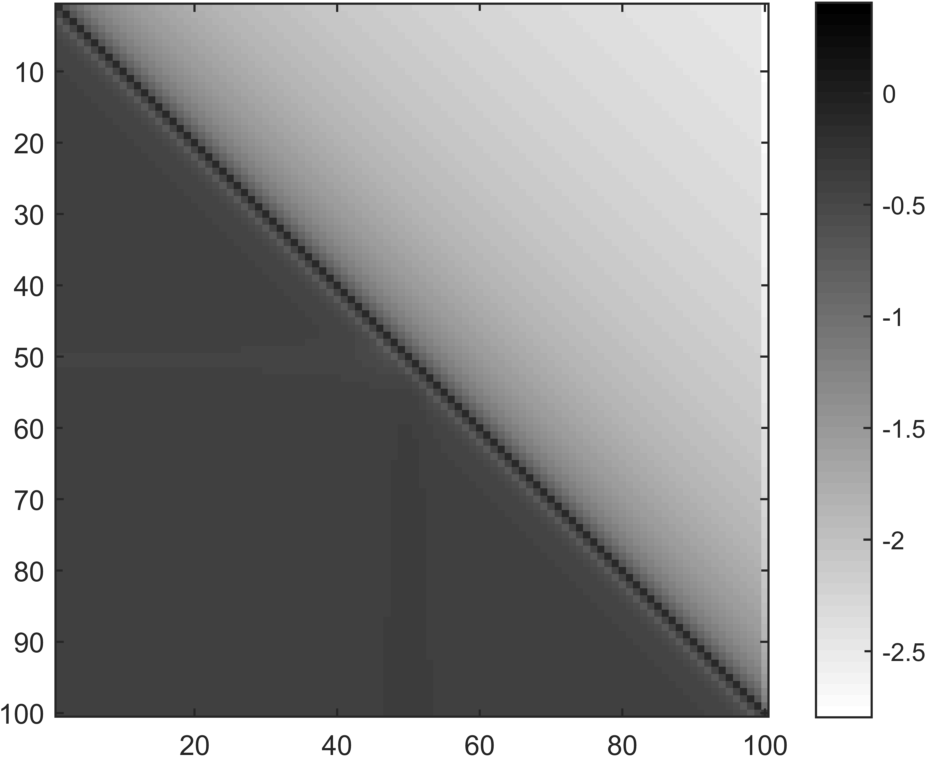}}
\caption{A visualisation of the magnitude of the (a) Jacobian and (b) Schur complement entries for the problem of flow past a pressure distribution given by (\ref{2Deq:pressure}) where $\textbf{u}_k=\textbf{0}$, $N=101$, $\phi_1=-20$, $\Delta\phi=0.4$, $\epsilon=0.1$ and $F_L=1$. The entries are on a log scale: $\mathrm{log}_{10}|J_{i,j}|$ for all $i$, $j$. Each element is assigned a shade based on its value: the larger the value, the darker the shade.}
\label{2Dfig:fullJac}
\end{figure}

A better approach exploits the structure of the Jacobian (Figure~\ref{2Dfig:fullJac}(a)), noticing that it can be partitioned into four submatrices:
\begin{equation}
\textbf{J}=\left[
\begin{matrix}
A & B\\
C & D\\
\end{matrix}\right],\label{2Deq:PreStruct}
\end{equation}
where $A$, $B$ and $C$ are lower bidiagonal and $D$ is fully dense. Block $LU$ factorisation of this matrix requires only the storage of $A$, a sparse matrix, and the Schur complement $D-CA^{-1}B$ (Figure~\ref{2Dfig:fullJac}(b)), while $B$ and $C$ need only be actioned as matrix products. This approach greatly reduces the storage requirements and factorisation time when using the Jacobian as the preconditioner matrix.

\begin{figure} 
\centering
\subfloat[Banded preconditioner]{\includegraphics[width=.46\linewidth]{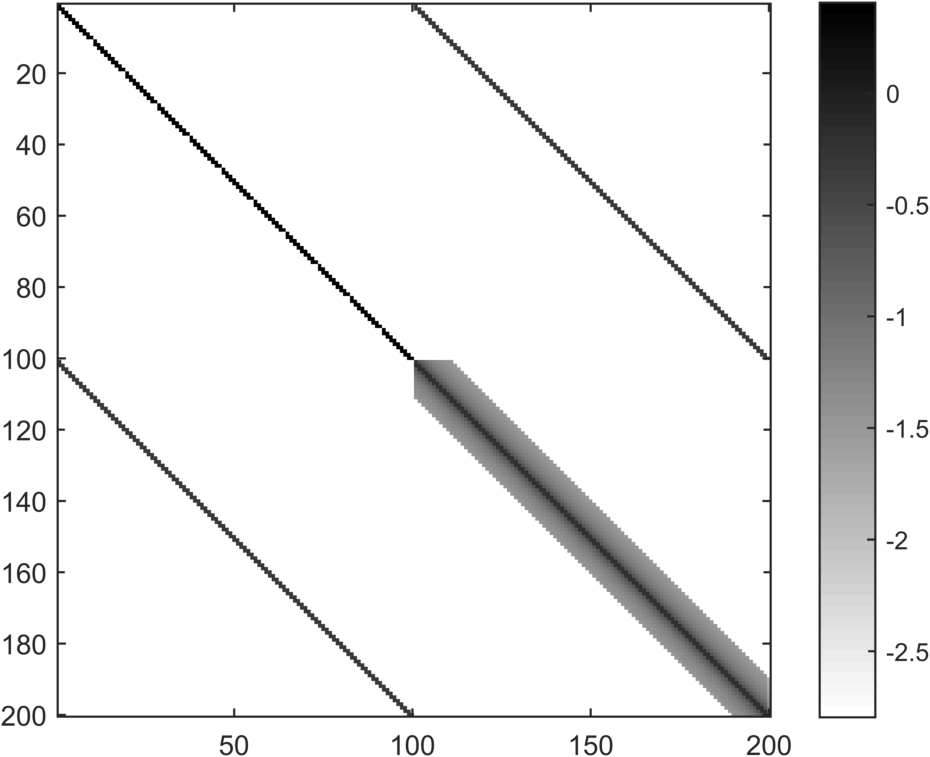}}\hspace{0.04\linewidth} 
\subfloat[Banded Schur complement]{\includegraphics[width=.46\linewidth]{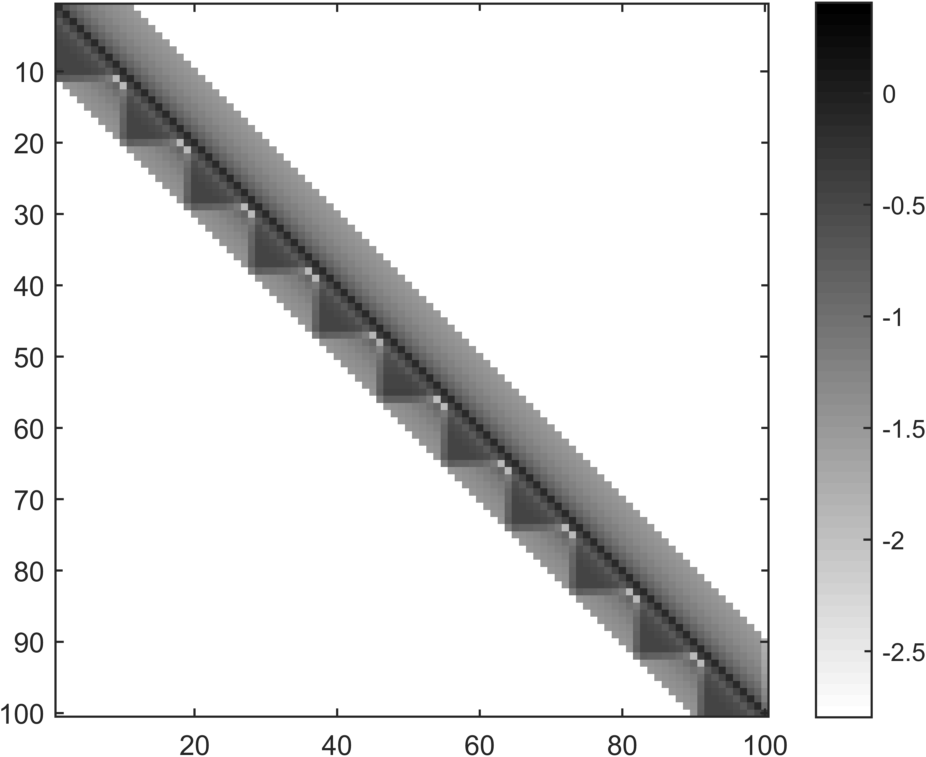}}
\caption{A visualisation of the magnitude of the (a) banded preconditioner and (b) banded Schur complement entries with upper and lower bandwidth $b=10$ for the problem of flow past a pressure distribution given by (\ref{2Deq:pressure}) where $\textbf{u}_k=\textbf{0}$, $N=101$, $\phi_1=-20$, $\Delta\phi=0.4$, $\epsilon=0.1$ and $F_L=1$. The entries are on a log scale: $\mathrm{log}_{10}|J_{i,j}|$ for all $i$, $j$. Each element is assigned a shade based on its value: the larger the value, the darker the shade.}
\label{2Dfig:bandJac}
\end{figure}

We can further reduce the storage and factorisation time by dropping selected matrix entries in order to force the Schur complement to be banded, thereby permitting banded storage and factorisation schemes to be used.  This approach is motivated by the observation in Figure~\ref{2Dfig:fullJac}(b) that the largest magnitude entries of the Schur complement occur within a small bandwidth of the main diagonal.  To retain values only within a bandwidth $2b+1$, we first set entries $A_{i+1,i}=0$ for $i=b+n(b-1)$ where $n=0,1,2,\dots$, thereby forcing $A^{-1}$ to have lower bandwidth $b-1$.  Then we set all entries of $D$ outside bandwidth $2b+1$ to be zero.  The computed Schur complement $D-CA^{-1}B$ then inherits the bandwidth $2b+1$.  A banded preconditioner and Schur compliment with upper and lower bandwidths $b = 10$ are shown in Figure~\ref{2Dfig:bandJac}.  Choosing an appropriate value of the bandwidth requires some experimentation, to find a suitable trade off between the efficiency of forming and factorising the preconditioner against its effectiveness at reducing the Krylov subspace dimension.

\subsection{Implementation}\label{2Dsubsec:implementation}
To implement the JFNK method, we used the Sundials KINSol package \cite{kinsol11} with a MATLAB interface.  The code to evaluate the nonlinear function $\textbf{F(\textbf{u})}$ was written in CUDA (Compute Unified Device Architecture) to run on a Graphics Processing Unit (GPU), which greatly reduces the evaluation time for this function. Factorising the preconditioner was performed using Intel's Math Kernel Library (MKL) to take advantage of a banded storage scheme. All remaining code, including forming the preconditioner as described in section \ref{2Dsubsec:Precon} and appendix~\ref{2Dsec:FormJac}, was written in MATLAB.  The high performance workstation used to produce the solutions in this paper included 2$\times$ Intel Xeon E5-2670 CPUs with 2.66 GHz processors, K40 Nvidia Tesla GPU and 124 GB of system memory.

To compute highly nonlinear solutions, a bootstrapping process was used, whereby a previously computed solution was used as the initial guess for the current solution. Typically, this previously computed solution was for a similar set of parameter values that corresponded to a less nonlinear version of the problem (eg. for flow past a pressure distribution, a previous solution with the same Froude number and a smaller value of $\epsilon$ was used as an initial guess).

\section{Infinite depth flow examples}\label{2Dsec:infinitedepth}

\subsection{Flow past a pressure distribution}
Flow past a pressure distribution has been well studied in both an infinitely deep fluid \cite{schwartz81,vandenbroeck84} or a finite depth channel \cite{grimshaw13,maki12,vandenbroeck96}.  A variety of variations to this problem have also been considered, such as the inclusion of the effects of surface tension on the free surface \cite{grimshaw13},  or even a layer of mud under the channel \cite{vandenbroeck96}.  However, for simplicity, we will only consider infinite depth flows without surface tension or other effects.

For definiteness, we consider flow past a pressure distribution which in dimensionless variables is given by
\begin{equation}
p(\phi) = \epsilon\mathrm{e}^{-\phi^2},\label{2Deq:pressure}
\end{equation}
where $\epsilon$ is the dimensionless pressure strength.  Since the velocity potential $\phi$ is monotonically increasing with $x$ (in a roughly linear fashion), the effect of (\ref{2Deq:pressure}) is that the pressure is localised about a particular point on the free surface and then decays off exponentially both far upstream (as $x\rightarrow -\infty$) and far downstream (as $x\rightarrow\infty$).
The governing equations are given by equations (\ref{2Deq:bernComplex}) and (\ref{2Deq:intAltered}) with $p(\phi)$ given by (\ref{2Deq:pressure}). The radiation condition for this problem (that the gravity waves can not develop in front of the disturbance) is enforced by $\tau_1=\theta_1=0$.

\begin{figure}
\centering
\subfloat[$\epsilon=0.05$]{\includegraphics[width=.8\linewidth]{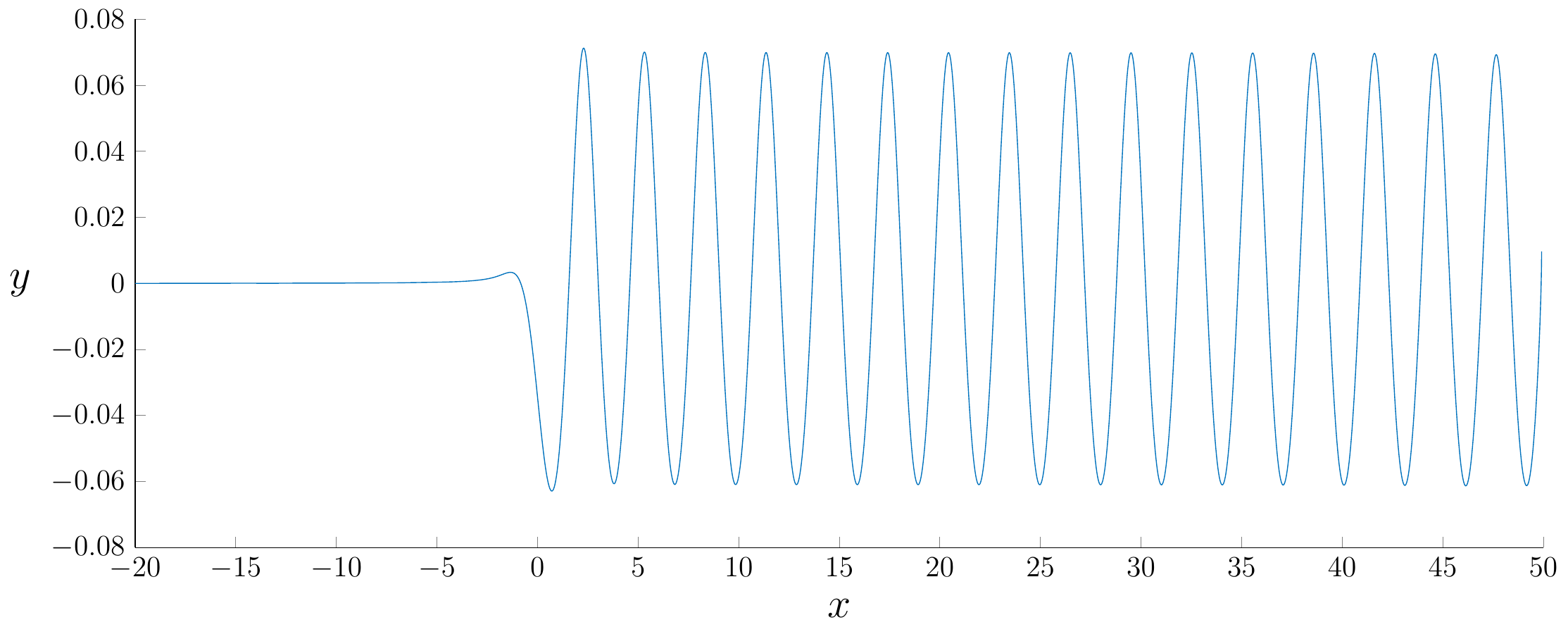}}\\
\subfloat[$\epsilon=0.1$]{\includegraphics[width=.8\linewidth]{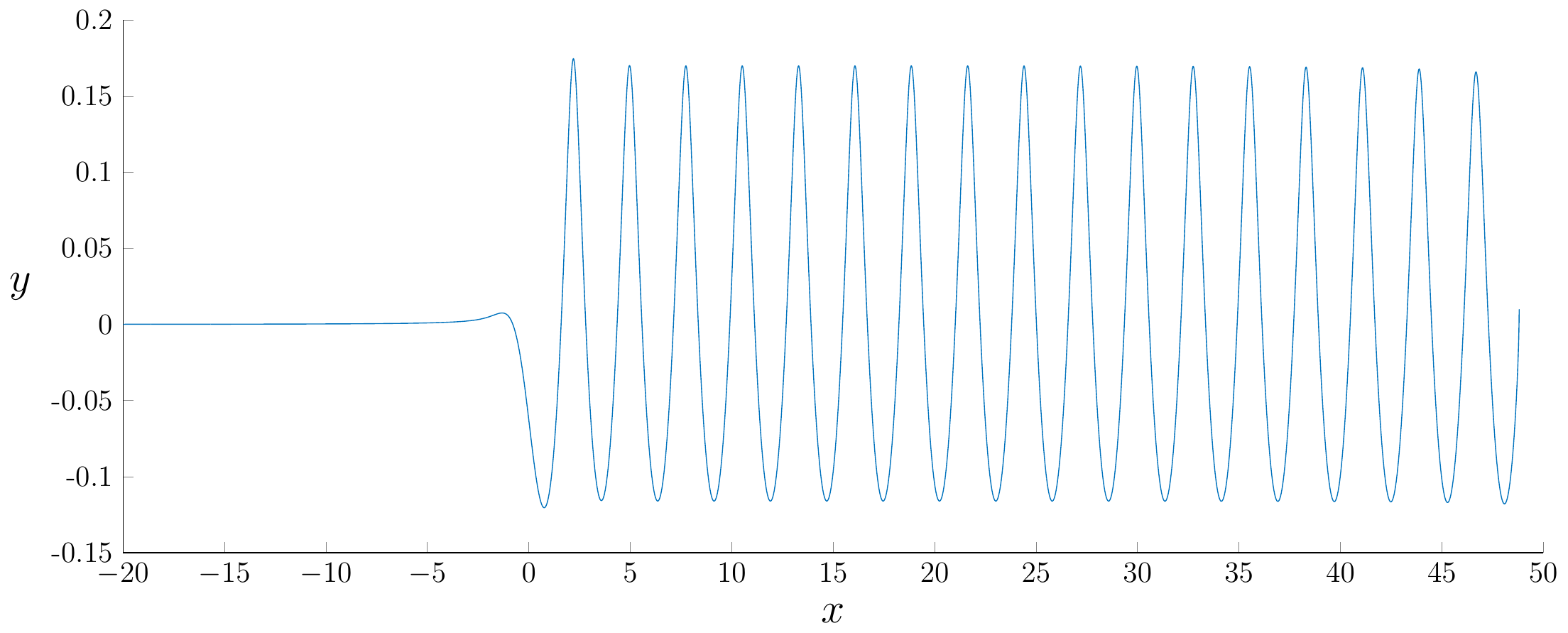}}
\caption{Free-surface profiles for flow past a pressure distribution (\ref{2Deq:pressure}) for $F_L=0.7$ with $N=120,001$, computed over the domain $-40<\phi<50$.  The steepness of the far field waves are: (a) $s=0.043$ for the pressure strength $\epsilon=0.05$; and (b) $s=0.103$ for $\epsilon=0.1$.}
\label{2Dfig:press}
\end{figure}

Using our numerical scheme, we are able to generate solutions on meshes with $N=120,001$ collocation nodes in under one hour on a high performance workstation.  A preconditioner matrix with a submatrix bandwidth of 48,001, occupying roughly 65GB of system memory, was used to compute these solutions. For example, Figure \ref{2Dfig:press} shows two free surface profiles for flow past the pressure distribution (\ref{2Deq:pressure}) with $F_L=0.7$, computed using this mesh and submatrix bandwidth.  The profile in Figure~\ref{2Dfig:press}(a) is for a moderately large pressure strength, $\epsilon=0.05$, while the profile in Figure~\ref{2Dfig:press}(b) is for a large pressure strength, $\epsilon=0.1$.

The solution in Figure~\ref{2Dfig:press}(b) represents a highly nonlinear solution.  A measure of this nonlinearity is that the maximum surface height is $\eta_{\mathrm{max}}=0.175$ which, for $F_L=0.7$, is roughly 71\% of $\eta_{\mathrm{upper}}=F_L^2/2$.  Another measure of nonlinearity is the steepness, $s$, of the periodic waves in the far field.  Here steepness is defined to be the wave amplitude divided by its wavelength.  The far field waves in Figure~\ref{2Dfig:press} have a steepness of roughly $s=0.1$, which is indicative of a highly nonlinear solution.  To put this into perspective, the absolute upper limit on the steepness is $s_{\mathrm{upper}}=0.141$, which corresponds to the Stokes limiting configuration characterised by $\eta_{\mathrm{max}}=\eta_{\mathrm{upper}}$ and a $120^\circ$ angle at the wave crest.  While there has been extensive research devoted to studying the shape of a single highly nonlinear gravity wave \cite{cokelet77,dallaston10,lukomsky02,schwartz74,williams81}, it is particularly challenging to compute solutions to full free surface flow problems (including the disturbance that causes the waves) with a train of waves with steepness of $s>0.1$.

The total time required to bootstrap up to the $\epsilon=0.1$ solution in Figure \ref{2Dfig:press} was approximately 35 hours. This approach with $N=120,001$ represents a substantial increase in the number of collocation points used for two-dimensional free-surface flow problems in recent times. For example, the number of mesh points typically used is roughly $500< N< 2000$ \cite{hocking13,holmes16,lustri12,trinh11,vandenbroeck02,wade14,wang17a,wang17}. The increased number of points can allow for a solution with much higher resolution or over a much larger domain.

In an attempt to quantify the benefits of using a large number of grid points, we perform a brief convergence study.  We generate solutions over seven different mesh sizes $N$ with the same computation domain in $\phi$ and three different pressure strengths $\epsilon$, remembering that the larger the value of $\epsilon$, the more nonlinear the solution.  We compare the steepness of the sixth wave (chosen to allow for the free-surface to approach its periodic behaviour) for each solution in Table~\ref{tab:pressSteep}. We see that for a small pressure strength $\epsilon=0.01$, all of the computed solutions exhibited the same steepness to four decimal places.  This information suggests that for near-linear solutions, a very large number of mesh points is not required.  However, as nonlinearity increases, the reported steepness does not appear to converge (to four decimal place accuracy) until $N\geq 30,001$ for $\epsilon=0.05$ and $N\geq 60,001$ for $\epsilon=0.1$.  In particular, we note that for the highly nonlinear solution, $\epsilon=0.1$, a typical mesh of say $N=1876$ fails to provide even three decimal place accuracy for the steepness.  Thus, Table \ref{tab:pressSteep} supports the assertion that for highly nonlinear regimes, a large number of mesh points may be required to provide accurate solutions.

\begin{table}
	\centering
	\begin{tabular}{|c|c|c|c|c|c|c|c|}
		\hline
		$N$	& 1876 & 3751 & 7501 & 15,001 & 30,001 & 60,001 & 120,001 \\
		\hline
		$\epsilon=0.01$ & 0.0082 & 0.0082 & 0.0082 & 0.0082 & 0.0082 & 0.0082 & 0.0082 \\
		\hline
		$\epsilon=0.05$ & 0.0429 & 0.0433 & 0.0431 & 0.0432 & 0.0433 & 0.0433 & 0.0433 \\
		\hline
		$\epsilon=0.1$ & 0.1003 & 0.1022 & 0.1022 & 0.1025 & 0.1026 & 0.1027 & 0.1027 \\
		\hline
	\end{tabular}
	\caption{The steepness of the sixth wave for flow past a pressure distribution for different mesh sizes, $N$, and nonlinearity, $\epsilon$. All solutions were computed over the same $\phi$ domain.}
	\label{tab:pressSteep}
\end{table}
\subsection{Flow under a semi-infinite plate}
The problem of flow past a semi-infinite plate placed on the free surface is a common prototype for approximating stern waves behind a two-dimensional ship in infinite depth \cite{farrow95,trinh11,vandenbroeck80,vandenbroeck78,vandenbroeck77} or finite depth \cite{mccue99,mccue02,mccue00}. The plate considered can either be horizontal for mathematical simplicity \cite{vandenbroeck80} or curved for a better representation of a ship's hull \cite{farrow95,ogilat11,trinh11}. For the following work, we only consider a horizontal plate in an infinitely deep fluid.  For our purposes, this flow configuration provides an interesting test case as there are semi-analytical results to compare with.  In particular, there is a prediction that solutions exists for $F_L>2.235$, with the $F_L=2.235$ solution having downstream waves with $s\approx 0.128$ \cite{vandenbroeck80}.  This is a particularly nonlinear solution.

We can take the numerical scheme from section \ref{2Dsec:NumSch} and apply it to the specific case of flow under a semi-infinite plate displaced a dimensional distance $L$ from the undisturbed level $y=0$ for $x\leq 0$ (this distance is unity in the dimensionless problem). The governing equations (\ref{2Deq:bernComplex}) and (\ref{2Deq:intAltered}) hold with $p(\phi)=0$ for $\phi>0$. The integral in equation (\ref{2Deq:intAltered}) can be updated with the knowledge that the fluid surface is flat under the plate:
\begin{equation}
\tau(\phi)=\frac{1}{\pi}\dashint_{0}^{\phi_N}\frac{\theta(\phi')-\theta(\phi)}{\phi'-\phi}\mathrm{d}\phi'+\frac{\theta(\phi)}{\pi}\ln\left|\frac{\phi_N-\phi}{\phi_1-\phi}\right|,\qquad 0<\phi<\infty.
\label{2Deq:intPlate}
\end{equation}
Finally, the boundary conditions at the edge of the plate at $x=0$ are
\begin{equation*}
\tau_1=\frac{1}{2}\ln\left(1\pm\frac{2}{F_L^2,}\right),\qquad\theta_1=0,
\end{equation*}
where the $\pm$ in $\tau_1$ is taken as `$+$' for lowering the plate below the $x$-axis or `$-$' for raising the plate above the $x$-axis. The remainder of the numerical method remains unchanged.

\begin{figure}
\centering
\includegraphics[width=.6\linewidth]{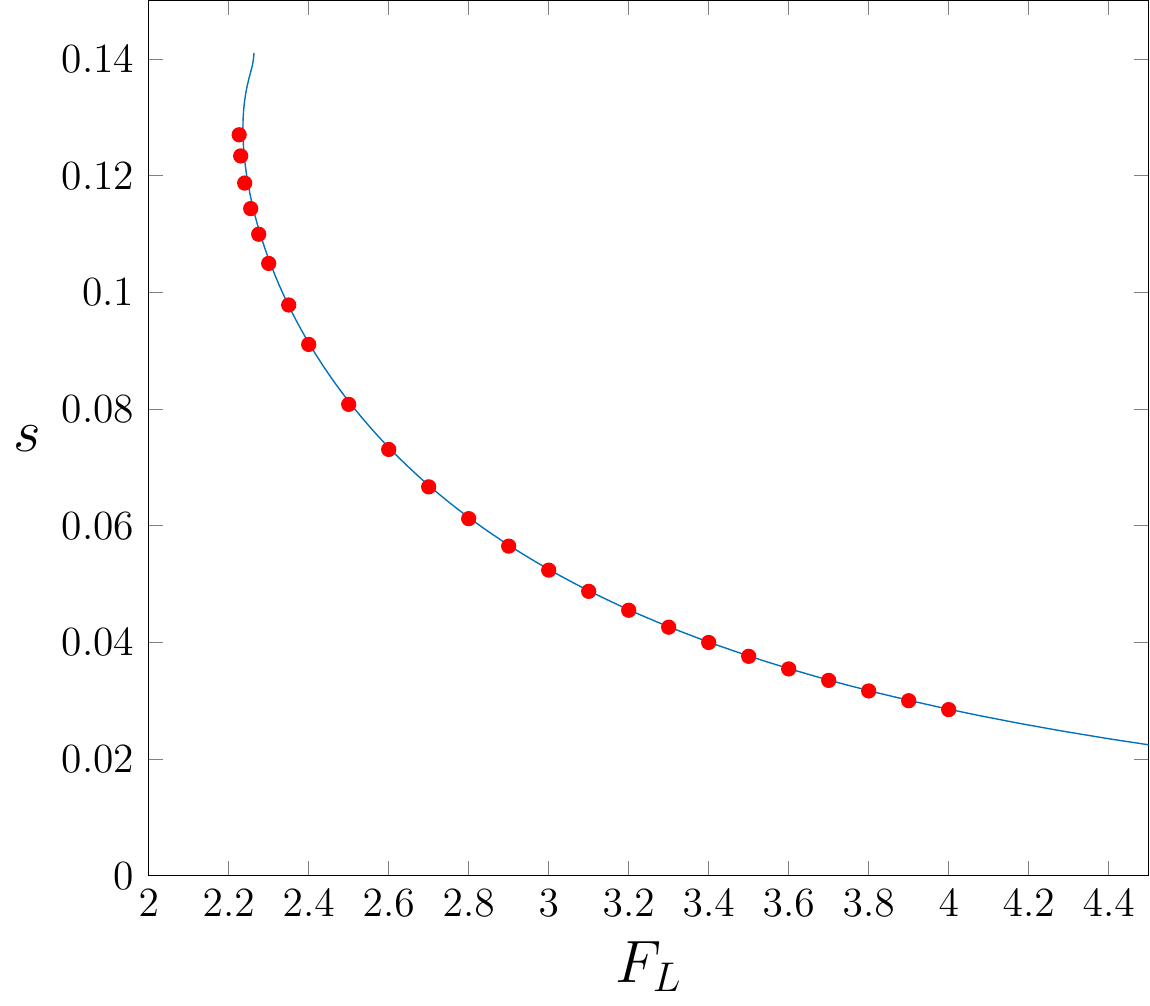}
\caption{A plot of the maximum steepness $s$ of a wave train for flow under a semi-infinite plate against the Froude number (red dots). The solid line is the semi-analytical relationship given by equation (2.2) in \cite{vandenbroeck80}, for which the curve is multivalued for roughly $2.235<F_L<2.264$.
}
\label{2Dfig:steepnessVsFr}
\end{figure}

As with flow past a pressure distribution, we use $N=120,001$ nodes and compute the solution using a preconditioner matrix with a submatrix bandwidth of 48,001. We have computed solutions for different Froude numbers and have plotted (with solid circles) the steepness of the waves in the far field for these solutions in Figure~\ref{2Dfig:steepnessVsFr}. Due to the large number of collocation nodes, we are able to compute solutions with very steep waves, while still maintaining high definition around the crest.  As a comparison, Vanden-Broeck~\cite{vandenbroeck80} derived a semi-analytical relationship, shown as the solid curve in Figure~\ref{2Dfig:steepnessVsFr}, between the wave steepness and the Froude number for this problem.  The relation predicts that the maximum steepness for a wave along the primary branch (the lower branch of the curve) is $s=0.128$.

\begin{figure}
\centering
\subfloat[$F_L=3$]{\includegraphics[width=.8\linewidth]{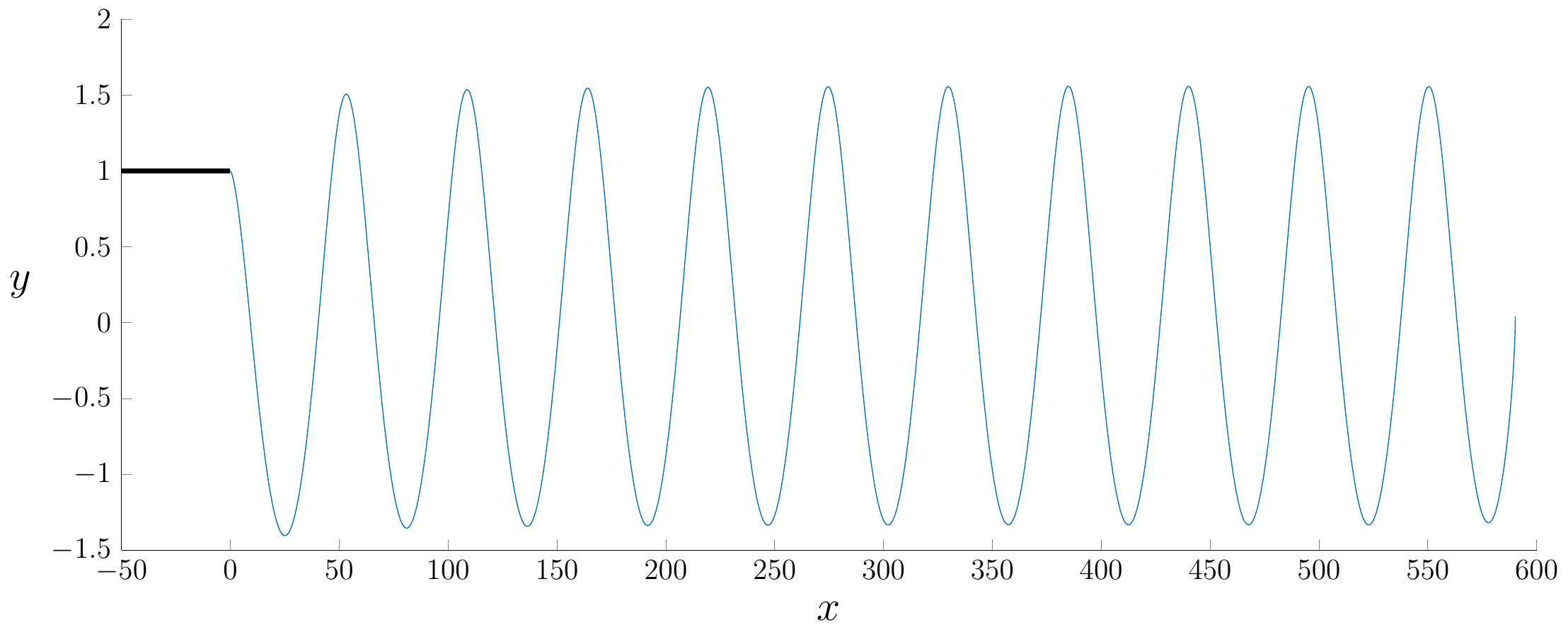}}\\
\subfloat[$F_L=2.22617$]{\includegraphics[width=.8\linewidth]{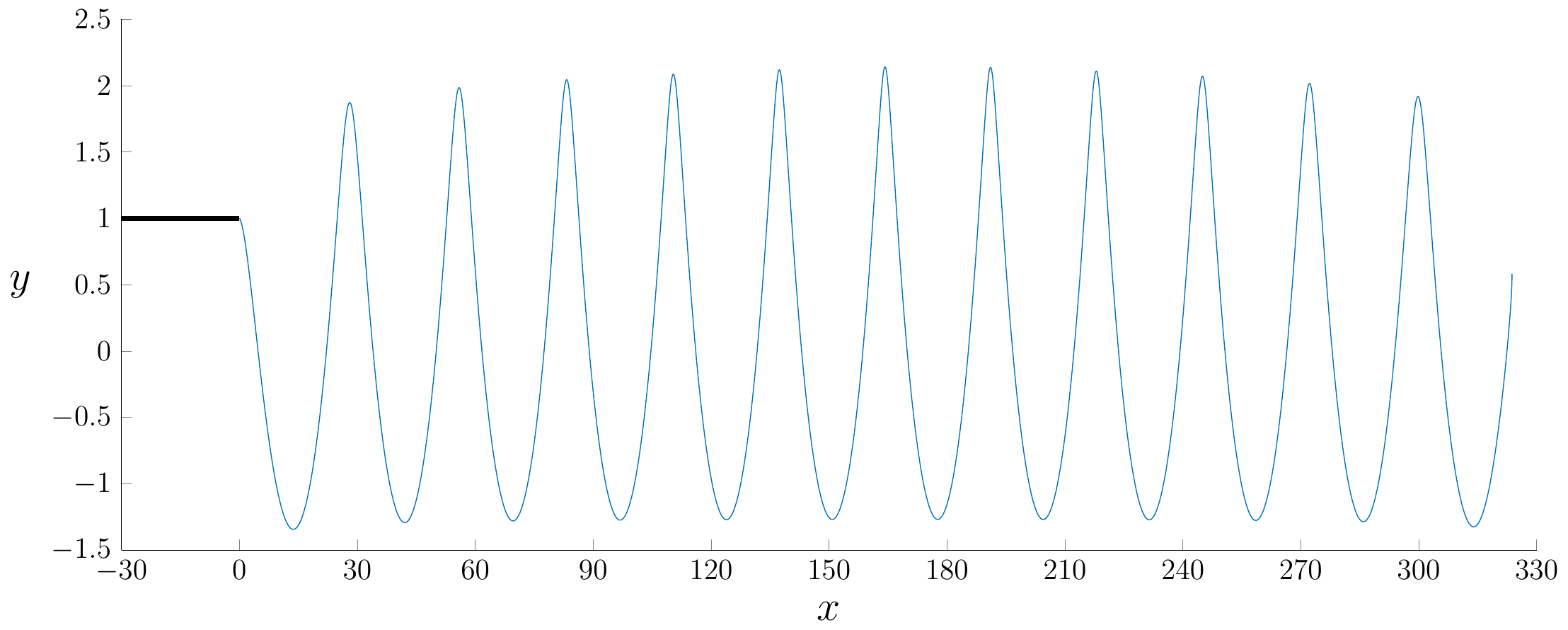}}
\caption{Free-surface profiles for flow under a raised semi-infinite plate for (a) $F_L=3$ and (b) $F_L=2.22617$ where $N=120,001$. The thick line represents the plate. The maximum steepness of the wave trains are (a) $s=0.0524$ and (b) $s=0.127$.}
\label{2Dfig:plate}
\end{figure}

Figure~\ref{2Dfig:plate} shows the free surface profiles for two examples, namely $F_L=3$ and $F_L=2.22617$.  The maximum steepness for the latter profile is $s=0.127$, which is close to the maximum possible steepness of $s=0.128$, indicating that this is a highly nonlinear solution.  An additional observation we can make from the two profiles in Figure~\ref{2Dfig:plate} is that for the less nonlinear solution, $F_L=3$, the crests and troughs of the waves quickly settle down to their far-field values.  On the other hand, for the highly nonlinear solution, $F_L=2.22617$, the crest heights do not appear to settle down, but instead increase and then decrease.  Thus, we see that, even with $N=120,001$ grid points, the truncation errors are affecting the wave profiles for highly nonlinear solutions in a manner that is observable by eye.

The results of a brief convergence study for this problem are presented in Table~\ref{tab:plateSteep}, where the maximum wave steepness is recorded for three different Froude numbers and many different grid sizes, remembering that for this problem, nonlinearity is increasing as the Froude number $F_L$ is decreasing.  We see that for the less nonlinear solutions ($F_L=5$ and $3$), there is little obvious gain in using a large number of grid points.  On the other hand, for the highly nonlinear solution with $F_L=2.23$, four decimal accuracy for the wave steepness is not achieved until roughly $N=30,001$.  Finally, for even lower values of $F_L$ such as the one used to compute the profile in Figure~\ref{2Dfig:plate}(b), four decimal place accuracy is not achieved even with $N=120,001$.

\begin{table}
	\centering
	\begin{tabular}{|c|c|c|c|c|c|c|c|}
		\hline
		$N$	& 1876 & 3751 & 7501 & 15001 & 30001 & 60001 & 120001 \\
		\hline
		$F_L=5$ & 0.0181 & 0.0181 & 0.0181 & 0.0181 & 0.0181 & 0.0181 & 0.0181 \\
		\hline
		$F_L=3$ & 0.0525 & 0.0524 & 0.0524 & 0.0524 & 0.0524 & 0.0524 & 0.0524 \\
		\hline
		$F_L=2.23$ & 0.1202 & 0.1223 & 0.1230 & 0.1232 & 0.1232 & 0.1232 & 0.1232 \\
		\hline
	\end{tabular}
	\caption{The maximum steepness of the wave train for flow under a semi-infinite plate for different mesh sizes, $N$, and Froude number, $F_L$. All solutions were computed over the same $\phi$ domain for a given Froude number.}\label{tab:plateSteep}
\end{table}

\section{Finite depth flow examples}
\label{2Dsec:finiteDepth}
A variant of the numerical scheme presented in Section~\ref{2Dsec:NumSch} can be used to compute finite-depth flows over a bottom topography. For finite-depth flows with an upstream dimensionless depth $\alpha$, dimensionless speed 1 and some bottom topography, we follow a similar procedure to the infinite depth case by mapping the physical $z$-plane to an infinite strip in the potential $w$-plane and then mapping the infinite strip to the half-$\zeta$-plane via $\zeta=\mathrm{e}^{\beta\pi w/\alpha}$, where $\zeta=\xi+i\eta$ and $\beta=\pm 1$. If $\beta=1$, the infinite strip is mapped to the lower half-plane with the upstream limit mapped to $\zeta=0$; otherwise, when $\beta=-1$ the infinite strip is mapped to the upper half-plane with the downstream limit mapped to $\zeta=0$. The choice of $\beta$ depends on the type of solutions computed. For subcritical flows, $F_H<1$, where $F_H=U/\sqrt{gH}$ is the depth-based Froude number, choosing $\beta=-1$ greatly reduces upstream numerical waves. Conversely, for supercritical flows, $F_H>1$, we choose to set $\beta=1$, to eliminate spurious numerical waves downstream. A schematic is shown for flow over a step with $\beta=-1$ in Figure~\ref{2Dfig:stepSchem}.

The integral equation (\ref{2Deq:intEq}) for finite depth flows is
\begin{equation}
\tau(\xi)=\frac{1}{\beta\pi}\dashint_{-\infty}^{\infty}\frac{\theta(\xi')}{\xi'-\xi}
\,\mathrm{d}\xi',\qquad-\infty<\xi<\infty.
\label{2Deq:intEqFinite}
\end{equation}
Typically, we know the function $\theta(\xi)$ for $\xi\in(-\infty,0)$ and can therefore break up the integral as
\begin{equation}
\tau(\xi)=f(\xi)+\frac{1}{\beta\pi}\dashint_{0}^{\infty}\frac{\theta(\xi')}{\xi'-\xi}\,\mathrm{d}\xi',\qquad 0<\xi<\infty,\label{2Deq:intEqFiniteBroken}
\end{equation}
where $f(\xi)$ is the integral evaluated over the known bottom. Finally, to implement the numerical scheme, we change our variable of integration back to $\phi$ and remove the singularity to give:
\begin{equation}
\tau(\phi)=f\left(\mathrm{e}^{\frac{\beta\pi\phi}{\alpha}}\right) +\frac{\beta}{\alpha}\int_{\phi_1}^{\phi_N}\frac{\theta(\phi')-\theta(\phi)}{1-\mathrm{e}^{\frac{\beta\pi}{\alpha}(\phi-\phi')}}\,\mathrm{d}\phi'+\frac{\theta(\phi)}{\pi}\ln\left|\frac{\mathrm{e}^{\frac{\beta\pi\phi_N}{\alpha}}-\mathrm{e}^{\frac{\beta\pi\phi}{\alpha}}}{\mathrm{e}^{\frac{\beta\pi\phi_1}{\alpha}}-\mathrm{e}^{\frac{\beta\pi\phi}{\alpha}}}\right|.
\label{2Deq:intFiniteAltered}
\end{equation}
The numerical scheme described in Section~\ref{2Dsec:NumSch} can now be applied to equations~(\ref{2Deq:bernComplex}) and (\ref{2Deq:intFiniteAltered}). Here, $F_L$ is replaced by $F_H$ where the reference length is the depth of the fluid far upstream, $H$.

It is worth noting a slight difference in the structure of the Jacobian for the finite-depth problems when compared to the infinite-depth problems.  The elements in the dense lower-right submatrix of the Jacobian for finite-depth problems no longer decrease to zero both above and below the diagonal. Instead, the elements approach $-\beta/\alpha$ away from the diagonal either above for $\beta=1$ or below for $\beta=-1$. Regardless, we implement the numerical scheme as before.

\subsection{Flow over a 90$\,^\circ$ step}\label{2Dsubsec:90step}
\begin{figure}
\centering
\subfloat[$z$-plane]{\includegraphics[width=.46\linewidth]{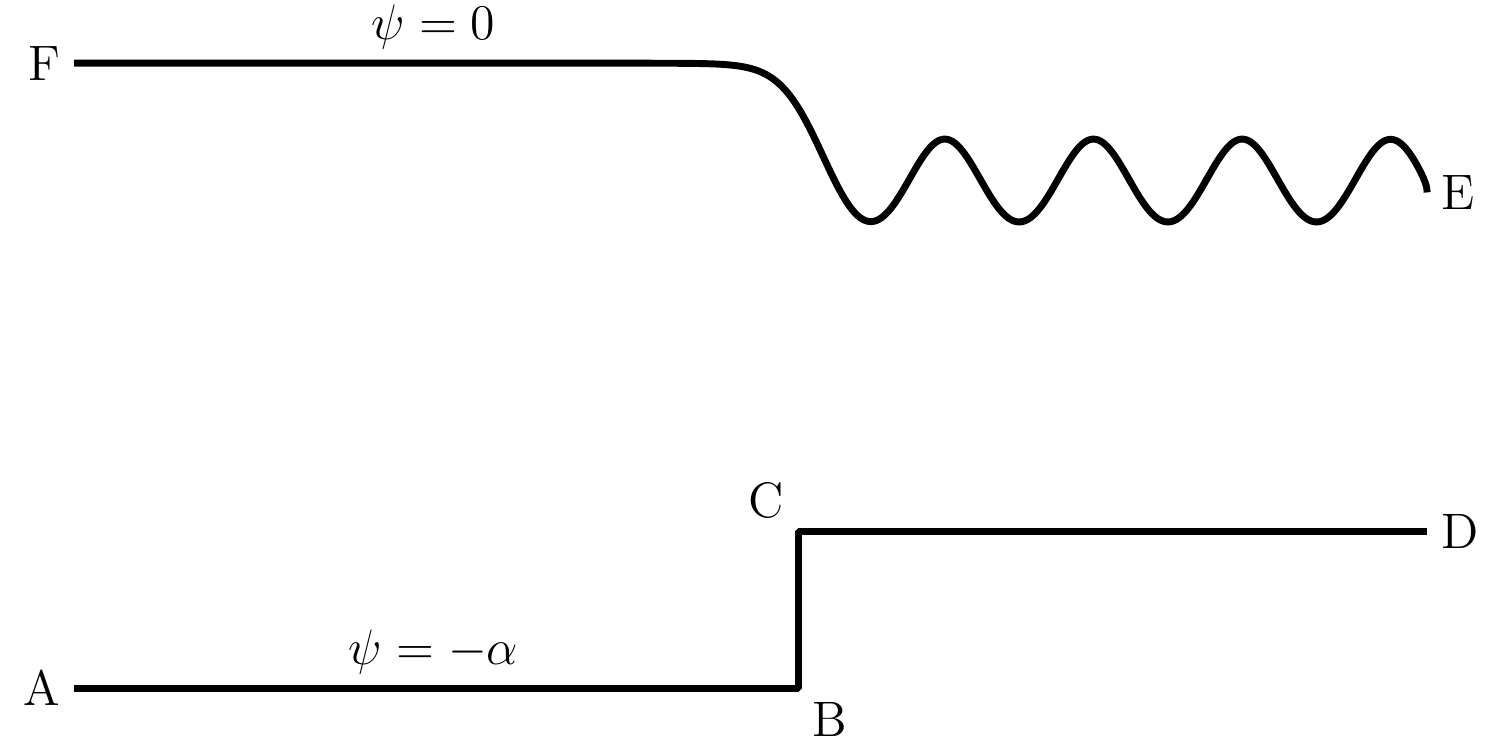}}\hspace{0.04\linewidth} 
\subfloat[$\zeta$-plane]{\includegraphics[width=.46\linewidth]{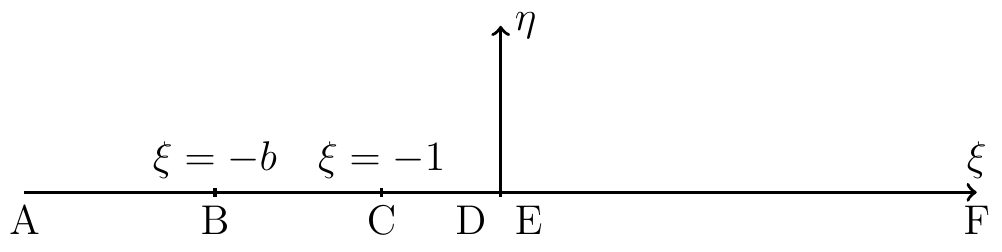}}
\caption{A schematic for flow over a step. The physical plane in part (a) is mapped to the upper half plane in part (b) via $\zeta=\mathrm{e}^{-\pi w/\alpha}$, where $w=\phi+i\psi$.}
\label{2Dfig:stepSchem}
\end{figure}

The problem of free-surface flow over a step has received considerable attention in the literature \cite{binder06,binder08,chapman06,king87,lustri12,toison00}.  Upstream and downstream from the step, the bottom topography is horizontal ($\theta=0$), while typically the step itself consists of a constant gradient segment connecting the two depths (see Figure \ref{2Dfig:stepSchem}(a) for a raised $90^\circ$ step). There are two different Froude numbers (upstream or downstream) that can be defined depending on which depth (before or after the step) is used. There are different flow regimes for flow over a step, depending on the properties of the flow.  We concentrate here on the subcritical regime where both the upstream and downstream Froude numbers are less than unity, for which a train of waves develops downstream from the step \cite{king87}.

We consider flow over a $90^\circ$ step as shown in Figure~\ref{2Dfig:stepSchem}, where we choose the $w$-plane such that the downstream edge of the step maps to $\xi=-1$ and the upstream edge maps to $\xi=-b$ where $\log(b)/\beta<0$. Here $b$ is a parameter related to the height of the step via (\ref{2Deq:stepH}) below. Thus the integral evaluated over the bottom, $f(\xi)$, is
\begin{equation}
f(\xi)=\frac{1}{2}\ln\left|\frac{\xi+b}{\xi+1}\right|\label{2Deq:StepEffect}
\end{equation}
and the radiation conditions are $\tau_1=\theta_1=0$. For the following results we choose $\alpha=\pi$ and $\beta=-1$.

To determine the height of the step, $h$, we integrate the identity $\mathrm{d}z/\mathrm{d}w=\mathrm{e}^{-\tau+i\theta}$ with respect to $\phi$ to give
\begin{equation}
h = \beta\int_{0}^{\frac{\alpha\log b}{\beta\pi}}\mathrm{exp}\left(-\tau\left(\phi-i\alpha\right)\right)\,\mathrm{d}\phi. \label{2Deq:stepH}
\end{equation}
The height of the step can be specified as an input to the problem by adding the equation (\ref{2Deq:stepH}) to the system of equations and the parameter $b$ to the vector of unknowns. The $2\times 2$ submatrix structure of the preconditioner (\ref{2Deq:PreStruct}) is then augmented with an additional row and column related to the equation (\ref{2Deq:stepH}) and the unknown $b$, respectively. The preconditioner can now be considered to have a $3\times 3$ submatrix structure. The additional 5 blocks, given by the additional row and column, are small in size and are kept in their entirety. Block $LU$ factorisation is applied the $3\times 3$ submatrix structure.

\begin{figure}[tb]
\centering
\subfloat[log(amplitude) vs. $b$]{\includegraphics[width=\linewidth]{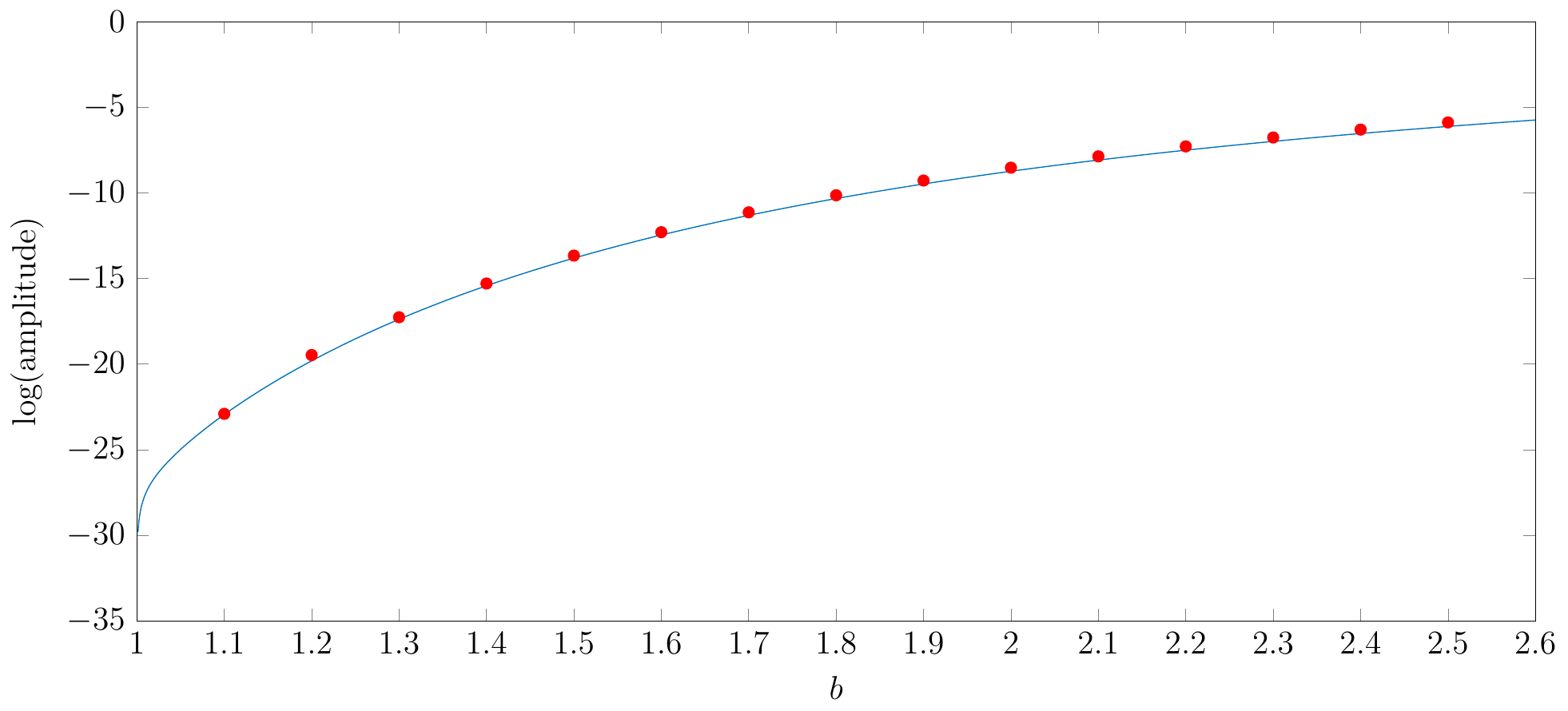}}\\
\subfloat[log(amplitude) vs. $F_H^2$]{\includegraphics[width=\linewidth]{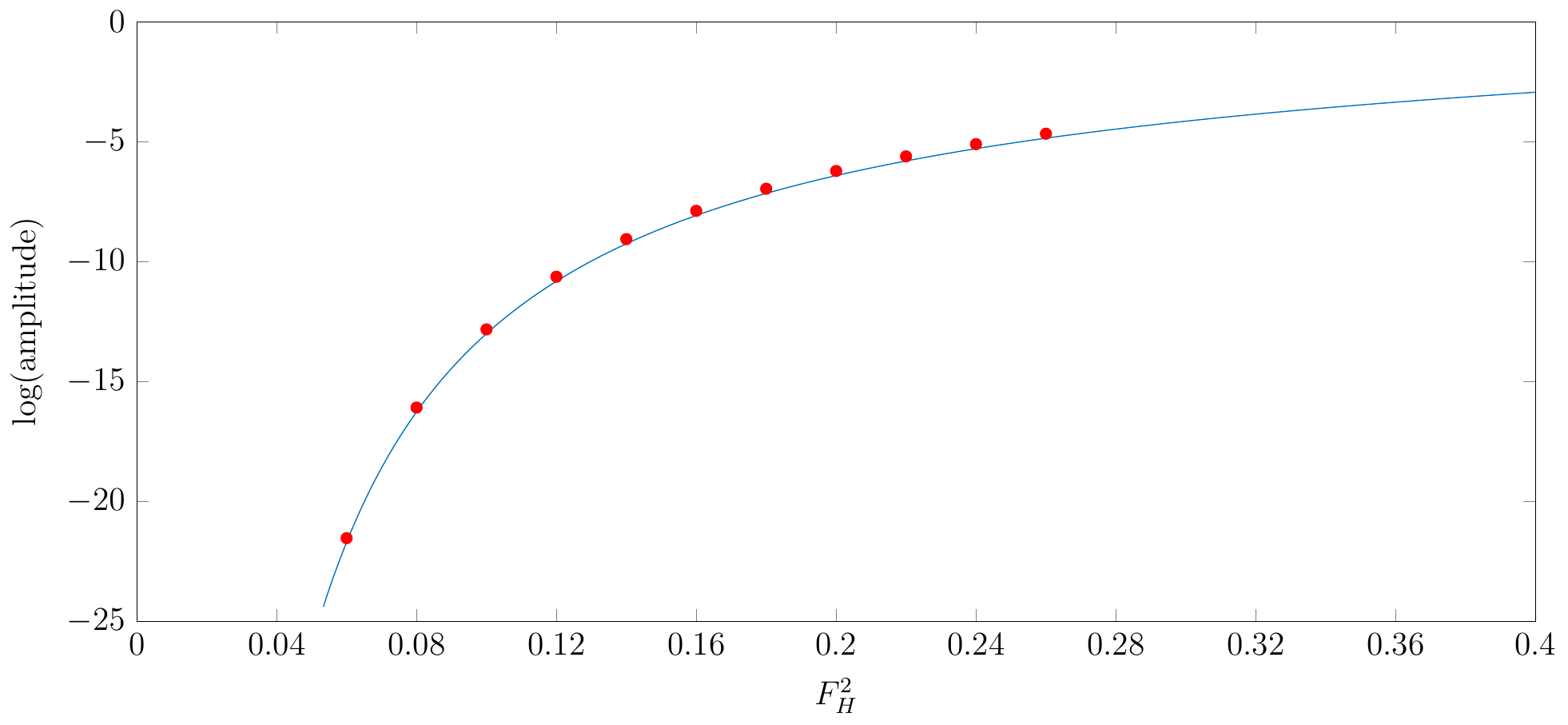}}
\caption{A plot of the log of the downstream wave amplitude against the parameter (a) $b$ with $F_H=0.2$ and (b) $F_H^2$ with $b=0.18$ for flow over a step . The solid curve is given in \cite{chapman06} as a uniform approximation generated from their equations (5.12), (B 16).}
\label{2Dfig:stepVsb}
\end{figure}

We use this flow configuration to illustrate how our numerical scheme can handle low Froude number regimes.  The challenge here is that the limit $F_H\rightarrow 0$ for fixed $b$ is singular \cite{chapman06,lustri12} and, furthermore, the amplitude and wavelength of the downstream waves becomes very small as the Froude number decreases (in fact the wavelength scales like $F_H^2$ and the amplitude scales like $F_H^\mu\mathrm{e}^{-\nu/F_H^2}$ as $F_H\rightarrow 0$, where $\mu$ and $\nu$ are positive constants).  Thus, for these reasons, a very large number of grid points is needed to resolve the free surface in the small Froude number limit.  As a test of our scheme, we reproduce the results of Chapman and Vanden-Broeck~\cite{chapman06} by analysing exponentially small waves generated by flow over a step for varying $b$ with constant Froude number (Figure~\ref{2Dfig:stepVsb}(a), which is similar to Figure 5 in \cite{chapman06}) and varying Froude number with constant $b$ (Figure~\ref{2Dfig:stepVsb}(b), which is similar to Figure 8 in \cite{chapman06}). In both cases, we demonstrate the ability to calculate solutions with exponentially small waves which is a computational challenge. We were able to resolve solutions with waves that are orders of magnitude smaller than those presented by Chapman and Vanden-Broeck~\cite{chapman06}. Additionally, the increased number of collocation nodes allows for sufficient resolution even on solutions with small wavelength (eg. roughly 165 nodes over a wavelength of 0.67 for $b=1.8$ and $F_H^2=0.06$).

\subsection{Flow over a curved step}
We now consider flow over a rounded step with a shape defined in the mapped plane by
\begin{equation}
\theta(\xi)=\begin{cases}
-\frac{\pi}{2(b-1)}(2\xi+b+1),&-b<\xi<-1\\
0,&\text{otherwise.}
\end{cases}
\label{eq:roundedstep}
\end{equation}
By evaluating the integral in (\ref{2Deq:intEqFinite}), we find the function $f(\xi)$ in (\ref{2Deq:intEqFiniteBroken}) is given by
\begin{equation}
f(\xi)=1+\frac{2\zeta+b+1}{2(b-1)}\log\left|\frac{\xi+1}{\xi+b}\right|.
\end{equation}
We apply the same numerical technique as with the $90^\circ$ step in subsection \ref{2Dsubsec:90step} to explore solutions for small Froude numbers $F_L\ll 1$.

\begin{figure}
\centering
\includegraphics[width=.6\linewidth]{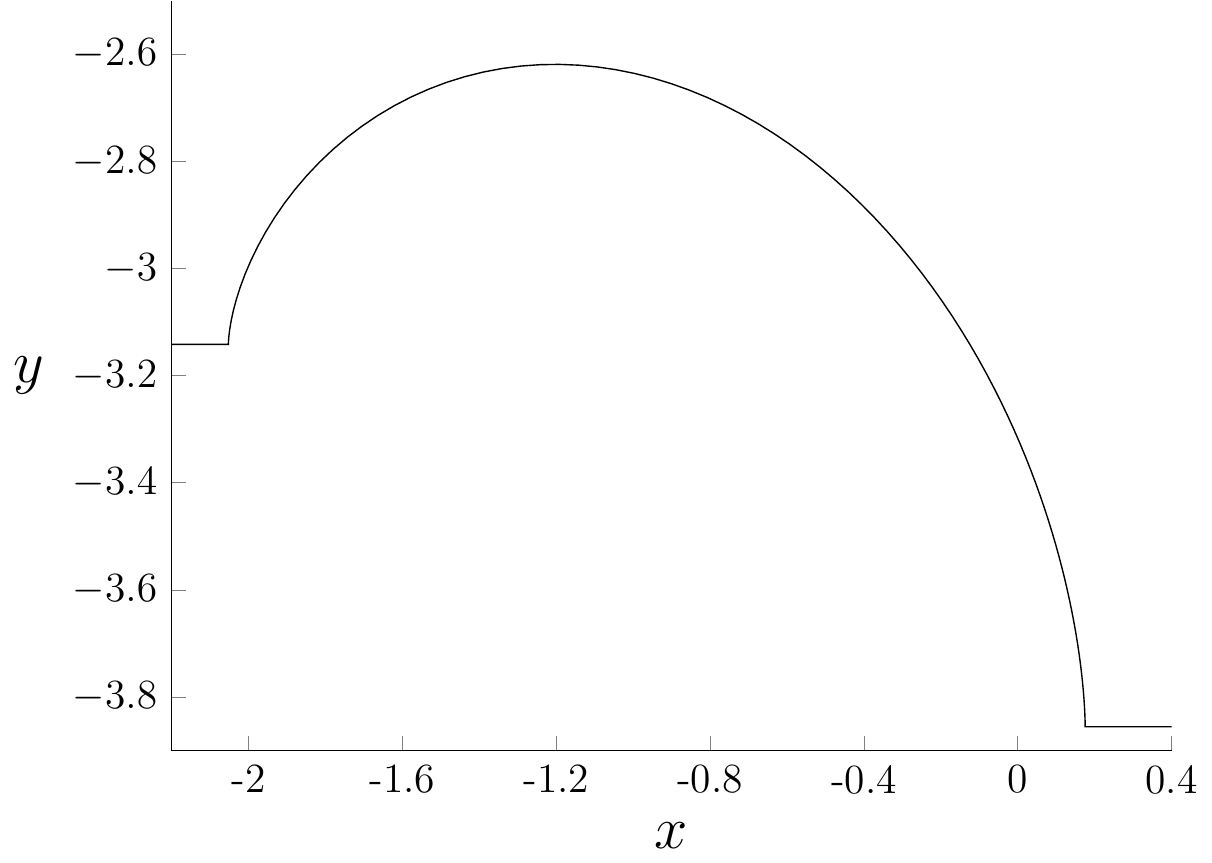}
\caption{An example of the hump shape for $b=6$, $F_H=0.6$ and $\alpha=\pi$.}
\label{2Dfig:humpExample}
\end{figure}

An example of the rounded step is shown in Figure~\ref{2Dfig:humpExample}.  This type of bottom configuration is worth considering because the theory of Chapman and Vanden-Broeck~\cite{chapman06} and Lustri et al.~\cite{lustri12} suggests that Stokes lines are generated by corners with in-fluid angles greater than $2\pi/3$, and that exponentially small waves are switched on where these Stokes lines intersect the free-surface (see also~\cite{trinh16,trinh14,trinh11}).  However, the rounded step (\ref{eq:roundedstep}) has two corners, both of which have in-fluid angles of $\pi/2$.  Thus there are no corners with in-fluid angles greater than $2\pi/3$ and, as such, a naive interpretation of \cite{chapman06,lustri12} is that there should not be waves generated by this bottom obstruction.

\begin{figure}[tb]
	\centering
	\subfloat[]{\includegraphics[width=.46\linewidth]{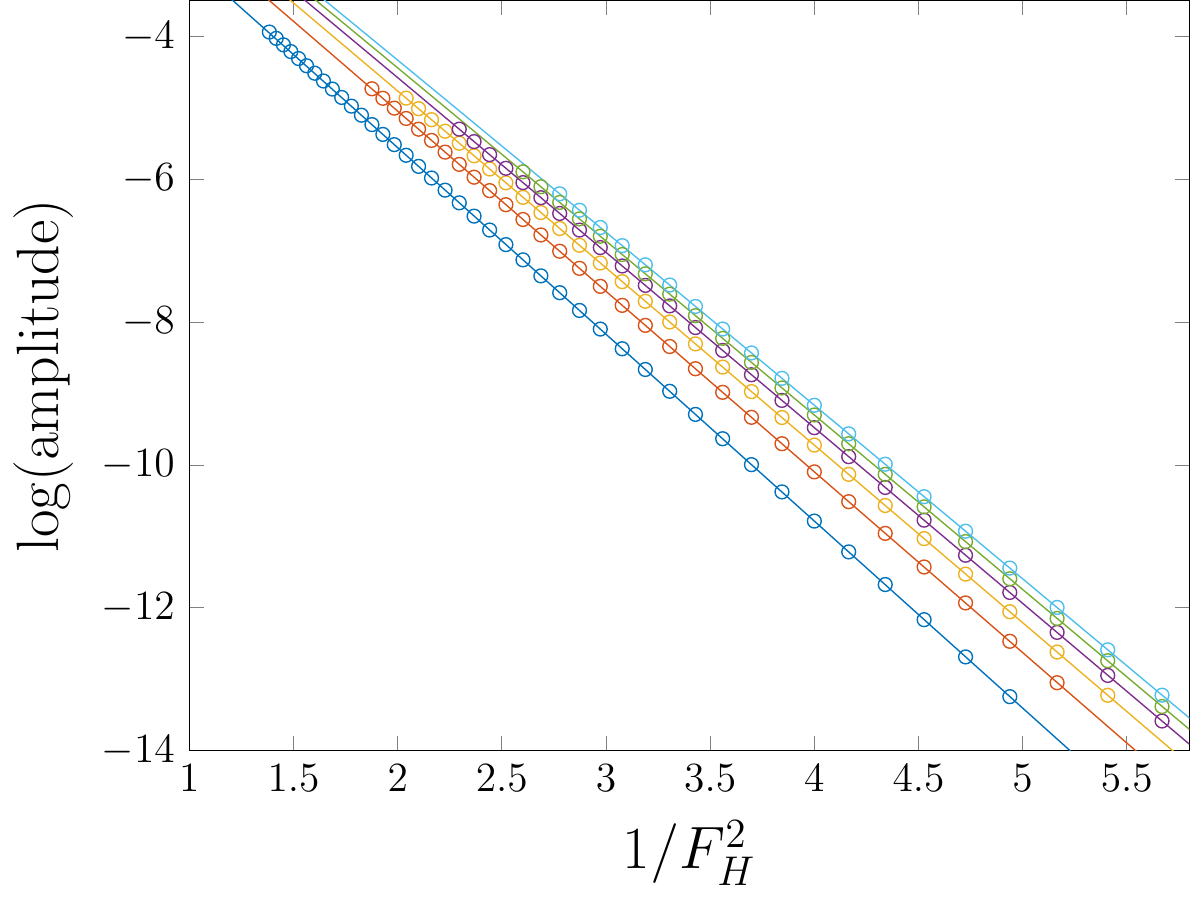}}\hspace{0.04\linewidth}
	\subfloat[]{\includegraphics[width=.46\linewidth]{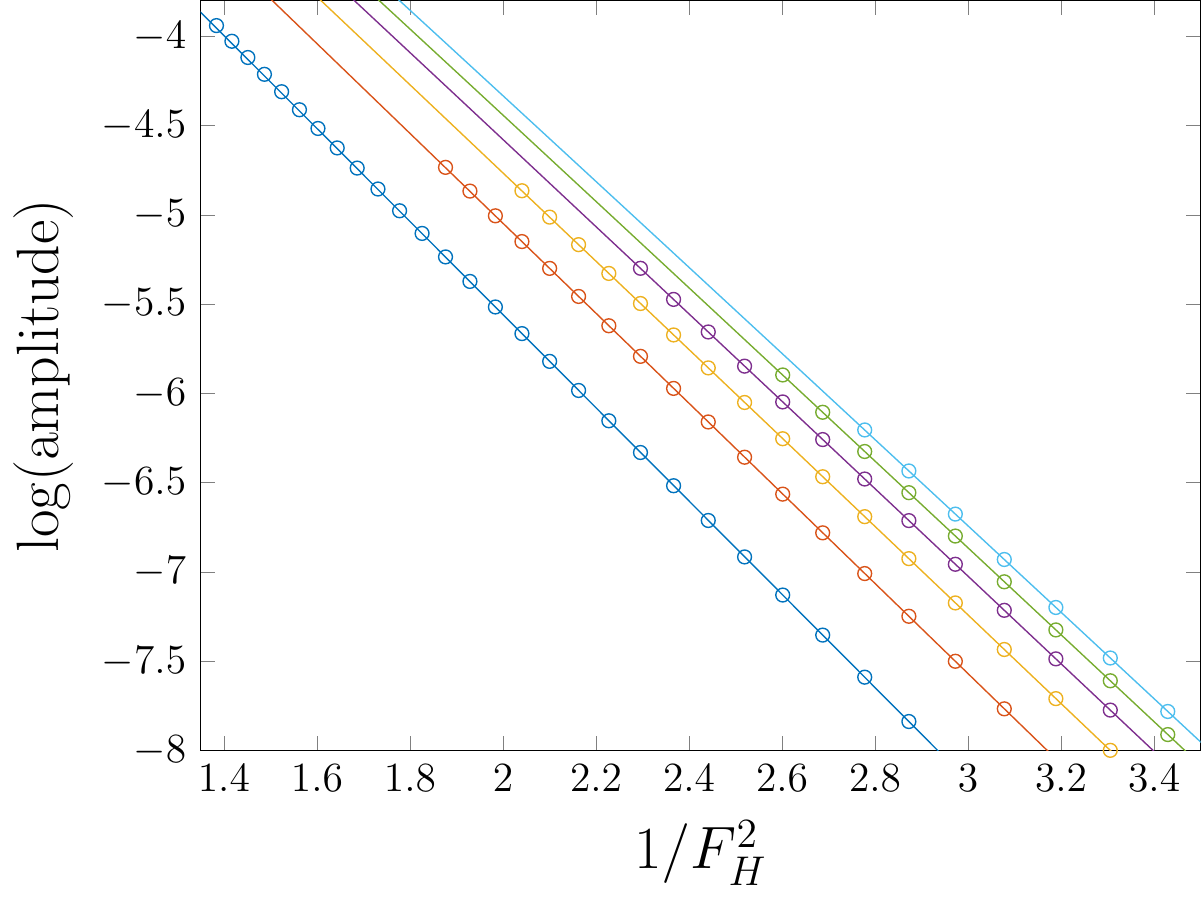}}
	\caption{A plot of the natural logarithm of the downstream wave amplitude for flow over a curved step against $1/F_H^2$. The circles are the data points computed from the nonlinear solutions to the curved step problem and the solid curves are the curves of best fit given by (\ref{2Deq:curveOfBestFit}) with the coefficients are given in Figure \ref{2Dfig:ceoffsVsb}. The results are presented for $b=3,4.5,\ldots,10.5$. (b) is a blown up version of (a).}
	\label{2Dfig:humpFittedData}
\end{figure}

\begin{figure}
	\centering
	\subfloat[$\log a$ vs $b$]{\includegraphics[width=.46\linewidth]{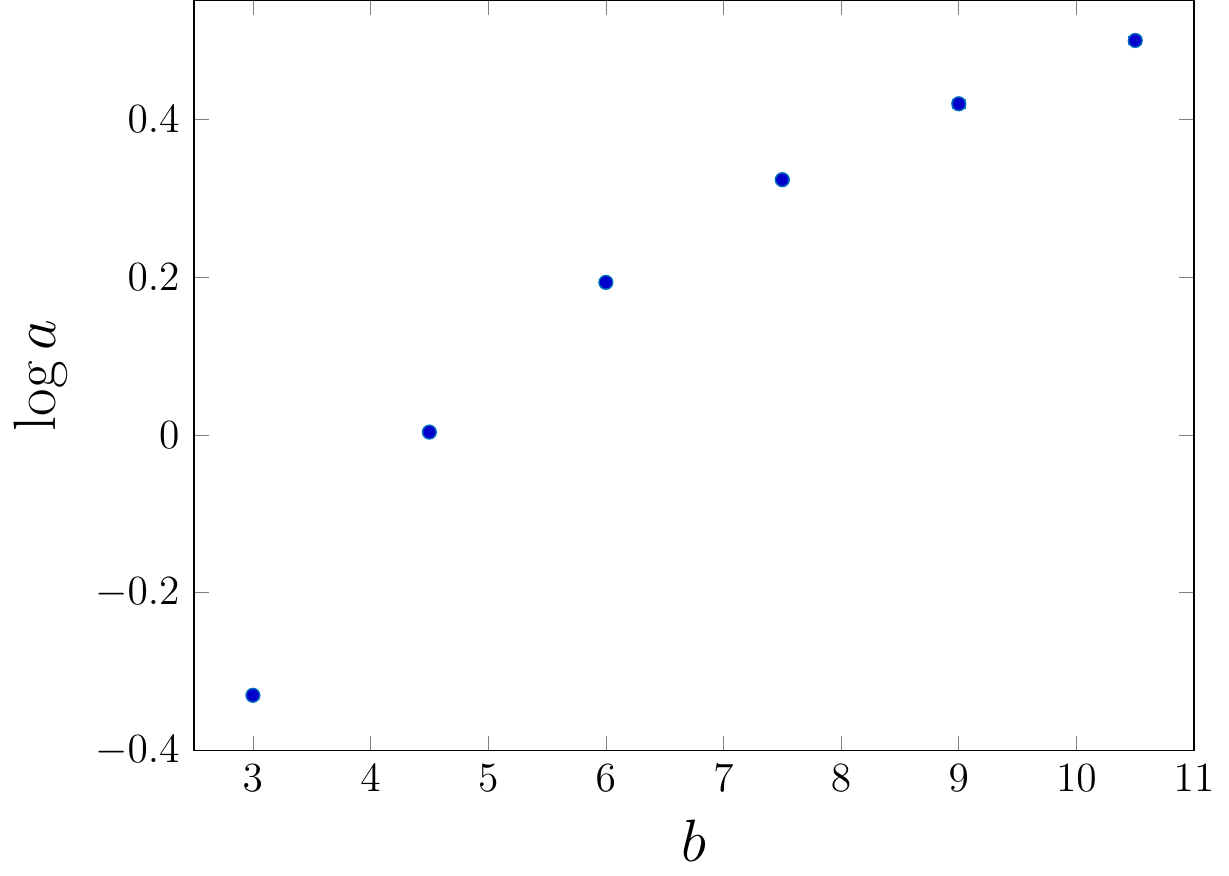}}\hspace{0.04\linewidth}
	\subfloat[$\mu$ vs $b$]{\includegraphics[width=.46\linewidth]{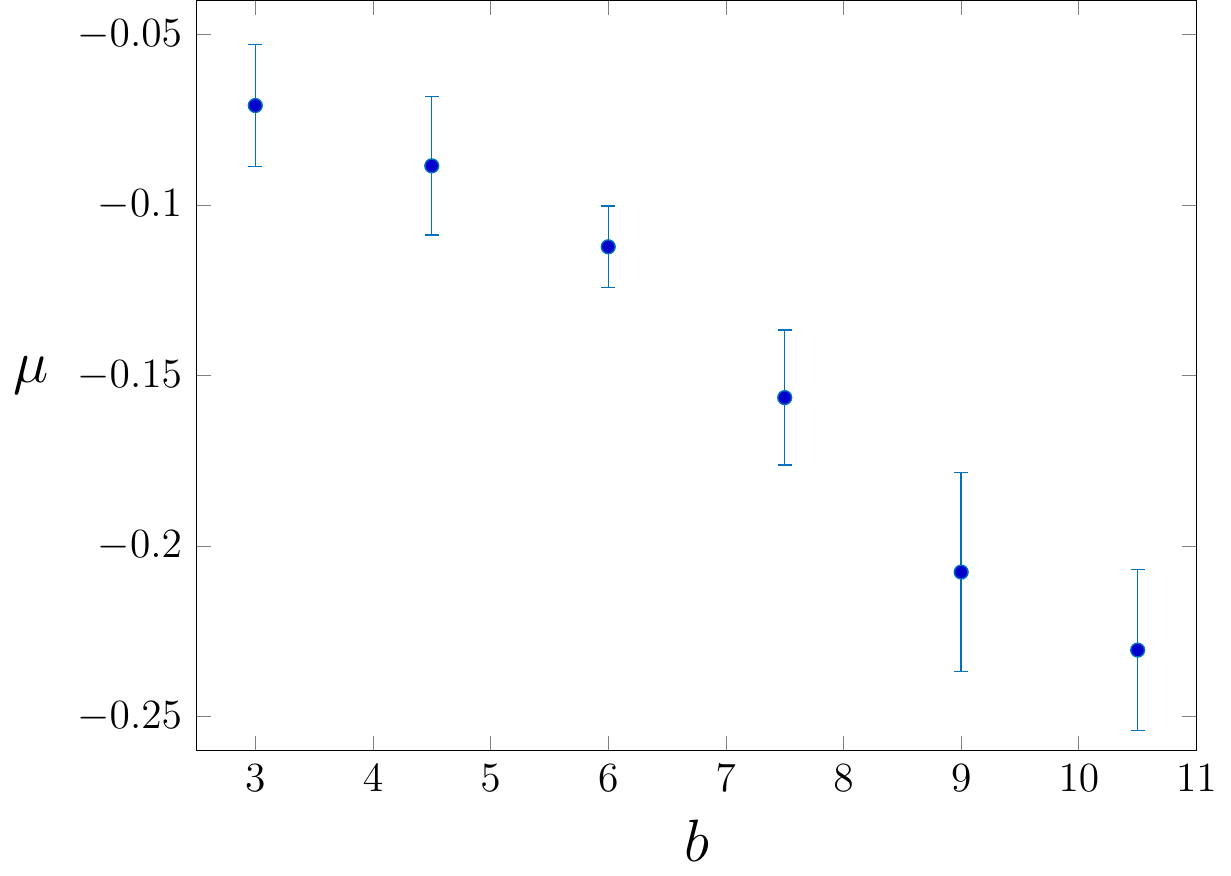}}\\
	\subfloat[$\nu$ vs $b$]{\includegraphics[width=.46\linewidth]{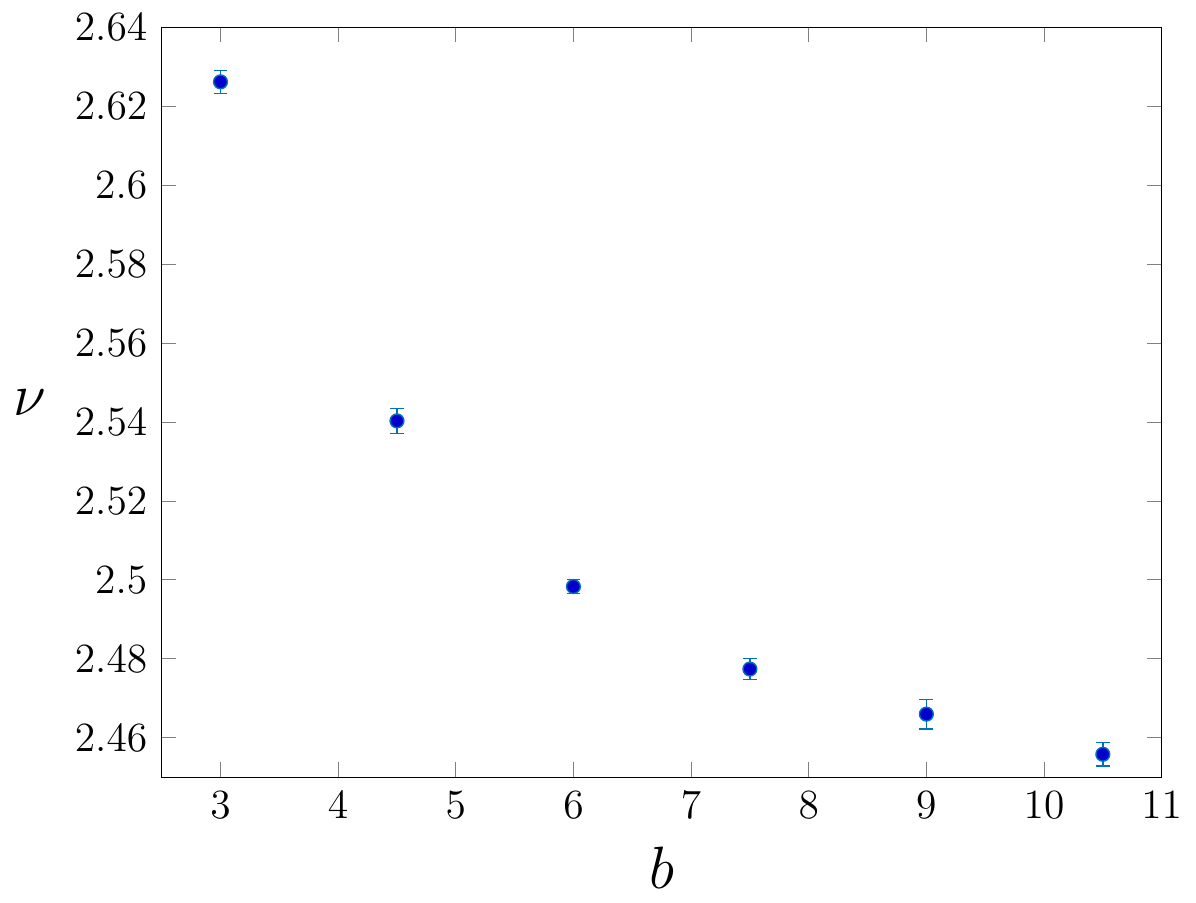}}
	\caption{A plot of the coefficients $a$, $\mu$ and $\nu$ of the curve of best fit (\ref{2Deq:curveOfBestFit}) along with their 95\% confidence interval {(ie. there is a 95\% probability the true coefficient values are within the interval)} for different values of $b$.}
	\label{2Dfig:ceoffsVsb}
\end{figure}

In contrast to this prediction, our numerical solutions do indeed show waves on the surface, and their amplitude appears to behave like $F_H^\mu\mathrm{e}^{-\nu/F_H^2}$ as $F_H\rightarrow 0$.  To demonstrate this behaviour, we have plotted the amplitude against the Froude number for constant values of $b$ and fitted a curve of the form
\begin{equation}
\text{amplitude}=aF_H^\mu\exp\left(-\frac{\nu}{F_H^2}\right),\label{2Deq:curveOfBestFit}
\end{equation}
to the data using a nonlinear least squares algorithm. When fitting the curves to the data, we are mindful of the range of data we consider because if the Froude number is too large, the approximation (\ref{2Deq:curveOfBestFit}) will no longer hold.  With this in mind, we only fitted data for which $-13<\log(\mathrm{amplitude})<-5$.  We have presented our results in Figure \ref{2Dfig:humpFittedData} for parameters $b=3,4.5,\ldots,10.5$.  In this figure, the open circles correspond to the numerical results while the solid curves are given by fitting (\ref{2Deq:curveOfBestFit}) to the numerical results for each value of $b$. The fitted coefficients $a$, $\mu$ and $\nu$ along with their 95\% confidence interval (ie. there is a 95\% probability the true coefficient values are within the interval) for different values of $b$  are presented in Figure \ref{2Dfig:ceoffsVsb}.  Thus we see for this example there are clearly waves on the free-surface which behave like (\ref{2Deq:curveOfBestFit}) in a way which is consistent with other problems with sharp corners.  The numerical results in Figure \ref{2Dfig:ceoffsVsb} should be used as a test for any future small-Froude-number asymptotics that does not require corners with in-fluid angles greater than $2\pi/3$.

To conclude, we have been able to use our numerical scheme to produce convincing numerical results that suggest there are indeed waves on the surface for flow past the rounded step (\ref{eq:roundedstep}) whose amplitude appears to scale like $F_H^\mu\mathrm{e}^{-\nu/F_H^2}$ as $F_H\rightarrow 0$.  These results suggest there {\em is} a Stokes line intersecting the free surface even though no Stokes lines originate from the two corners in this configuration.  As such, there must be a Stokes line structure that is not described by Chapman and Vanden-Broeck~\cite{chapman06} or elsewhere, leaving an interesting open problem for further research in the area of exponential asymptotics.

\subsection{Flow over a triangular obstruction}
Another finite depth configuration that has been extensively studied is flow over an obstruction where the averaged depth far upstream and far downstream is the same \cite{binder13,binder05,cole83,dias89,forbes82,forbes88,lee15,lustri12,toison00,vandenbroeck87,zhang96}.
Many different obstructions have been explored: triangles \cite{binder05,dias89}, semicircles \cite{forbes88,forbes82,vandenbroeck87}, or an arbitrary curved surface \cite{toison00}. Like flow over a step, flow over a obstruction can be classified as subcritical, exhibiting downstream waves, or supercritical, exhibiting no waves.  We choose here to focus on the supercritical regime because there is evidence of nonuniqueness that can only be resolved with a particularly fine mesh.  This is therefore a good test for our numerical scheme.

We will consider flow over an isosceles triangular obstruction, because it can be simply defined by two flat surfaces with an internal angle to the horizontal and a height. The function, $f(\xi)$, can be evaluated, giving
\begin{equation}
f(\xi)=\frac{\Theta_I}{\pi}\ln\left|\frac{(\xi+b)(\xi+a)}{(\xi+1)^2}\right|,\label{2Deq:TriEffect}
\end{equation}
where $\Theta_I$ is the interior base angles of an isosceles triangle, $\log(a)/\beta>0$ is the downstream edge of the triangle in the $\zeta$-plane and $b$ is the same parameter as with flow over a step. In the  results presented here we choose $\Theta_I=\pi/4$, $\alpha=1$ and $\beta=1$.

\begin{figure}
\centering
\includegraphics[width=.8\linewidth]{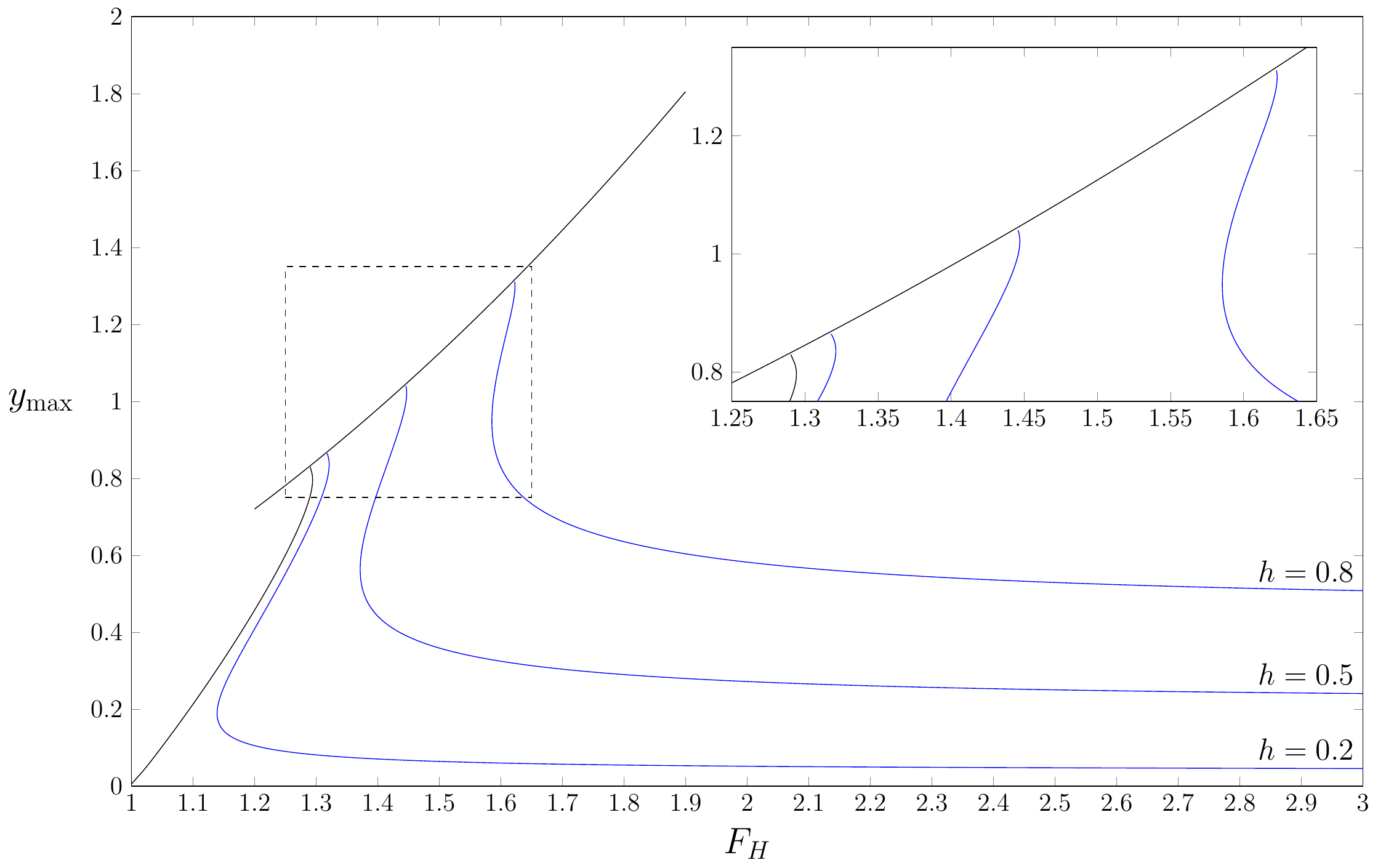}
\caption{A plot of the maximum wave height against the Froude number for constant values of triangle height, $h$. The black curves are the limiting configurations for supercritical flow over an obstacle.  An inset is given to more clearly show the second fold.}
\label{2Dfig:ymaxVsH}

\end{figure}

\begin{figure}
\centering
\includegraphics[width=.8\linewidth]{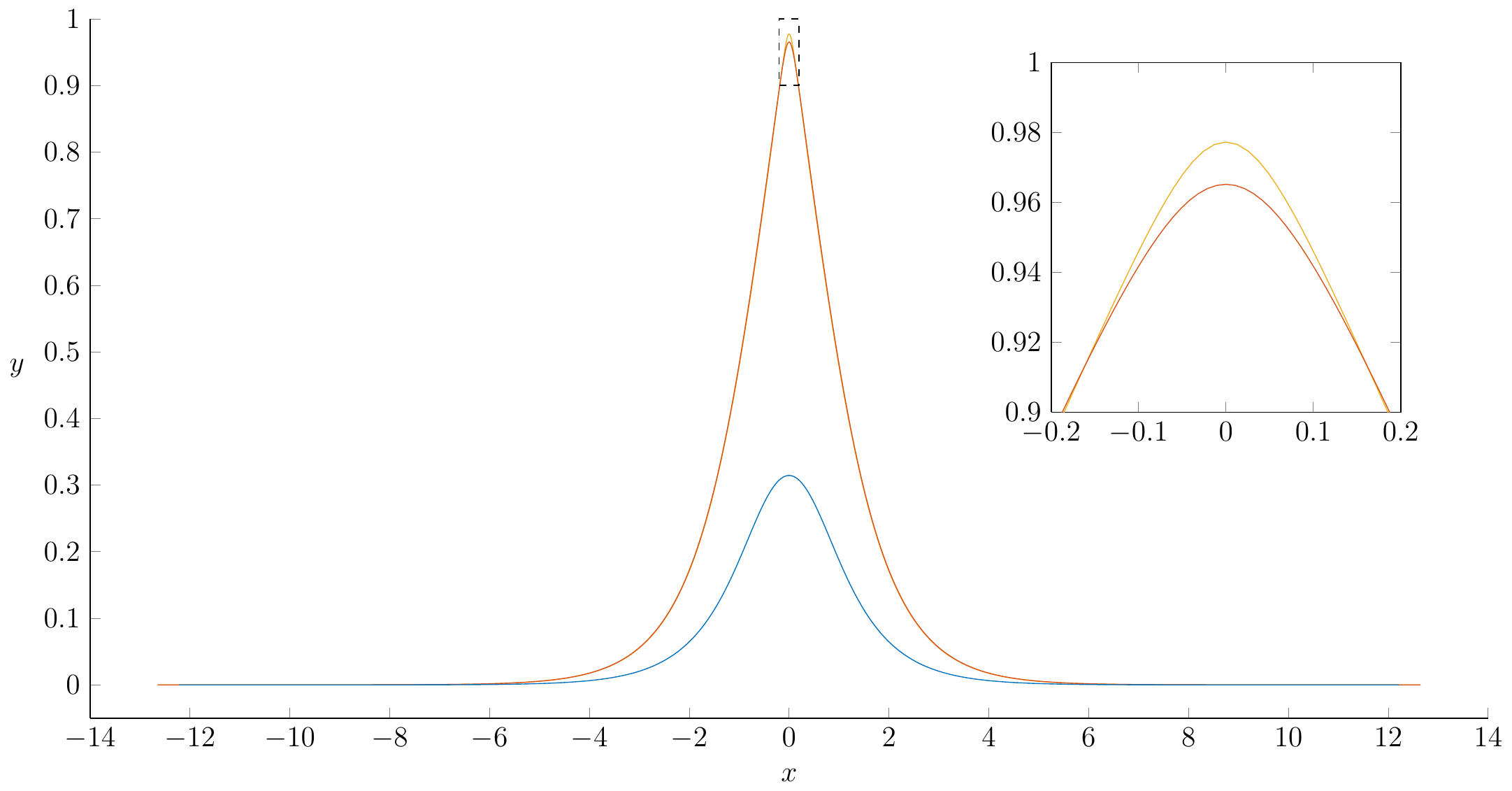}
\caption{The three possible free-surface profiles for supercritical flow over a triangular obstacle of height $h=0.437$ and $F_H=\sqrt{2}$.}
\label{2Dfig:3tri}
\end{figure}

We can see in Figure~\ref{2Dfig:ymaxVsH} that for a given triangle height $h$, there exists a Froude number, $F_H^*$, for which solutions only exist for $F_H>F_H^*$. If $y^*=\eta_\mathrm{max}(F_H^*)$, where $\eta_\mathrm{max}$ is the maximum height of the free surface, then it is known that the solutions where $\eta_\mathrm{max}<y^*$ are perturbations of the free stream $y=0$ and solutions where $\eta_\mathrm{max}>y^*$ are perturbations of the solitary wave solution \cite{wade14}. However,  it is possible to have three or possibly more solutions for a single value of triangle height, $h$, and Froude number.  An example is given in Figure~\ref{2Dfig:3tri}. This is due to the fact that the maximum solitary wave height is a multivalued function of the Froude number \cite{longuethiggins96,wade14} and thus multiple solutions can be found for a given Froude number by perturbing solitary wave solutions for above and below the fold. An example of multiple (three) solutions for a given triangle height, $h=0.437$, and Froude number, $F_H=\sqrt{2}$, is given in Figure~\ref{2Dfig:3tri}.

{ We close with a very brief convergence study for this problem.  In Table~\ref{tab:triHeight} we present the maximum surface height computed for three different solutions, each with a fixed obstruction height $h=0.5$.  The solution for $F_H=3$ is very close to the linear regime and does not require a large number of grid points to achieve a converged maximum surface height value.  The solution for $F=1.44$ on the lower branch is more nonlinear, but still does not require a large value of $N$ to obtain accurate computations.  On the other hand, the solution for $F=1.44$ on the upper branch is highly nonlinear.  In this case, we are not able to demonstrate four decimal place accuracy even with $N=120,001$ points.}


\begin{table}
	\centering
	\begin{tabular}{|c|c|c|c|c|c|c|}
		\hline
		$N$	& 3751 & 7501 & 15001 & 30001 & 60001 & 120001 \\
		\hline
		$F_H=3$	& 0.24046 & 0.24046 & 0.24046 & 0.24045 & 0.24046 & 0.24046 \\
		\hline
		$F_H=1.44$, lower branch& 0.39576 & 0.39576 & 0.39576 & 0.395756 & 0.39576 & 0.39576 \\
		\hline
		$F_H=1.44$, upper branch& 0.95616 & 0.95614 & 0.95615 & 0.95586 & 0.95616 & 0.95608
		\\
		\hline
	\end{tabular}
	\caption{The maximum surface height for flow over a triangular obstruction of height $h=0.5$ for different mesh sizes, $N$, and Froude number, $F_H$. All solutions were computed over the same domain.}\label{tab:triHeight}
\end{table}

\section{Discussion}
\label{2Dsec:Discussion}
\subsection{Options for implementing JFNK scheme}

We have developed a framework for applying a preconditioned Jacobian-free Newton-Krylov (JFNK) method to solve two-dimensional steady free-surface flow problems. Based on our experience, there are three levels of sophistication that can be used:
\begin{enumerate}
\item Implement the JFNK method using KINSol (or equivalent) with a preconditioner matrix given by the full Jacobian calculated as in appendix~\ref{2Dsec:FormJac} (or by finite-difference if the entries of the Jacobian cannot be determined exactly, although this can lead to infeasibly long runtimes for large mesh sizes). This is the most straightforward option. The KINSol package is simple to install and use and, importantly, its JFNK implementation is much more efficient than using the full Newton's method.
\item Perform implementation 1 while exploiting the structure of the Jacobian as in section \ref{2Dsubsec:Precon} to construct a preconditioner that makes more efficient use of system memory. Additionally, the use of Intel MKL (or equivalent) to performed banded factorisation of the Jacobian is recommended.  This level of implementation is the most efficient option for a standard desktop computer (without a GPU).
\item If a Graphics Processing Unit (GPU) is available, the evaluation of the nonlinear function $\textbf{F}(\textbf{u})$ can be coded in CUDA (or equivalent) to run in parallel on a GPU. Utilising a GPU will greatly decrease the runtime of either implementation 1 or 2.
\end{enumerate}
The implementation chosen may depend on the available hardware and the familiarity of the user with different programming languages and architectures.

\subsection{Versatility of our JFNK scheme}

Our approach is motivated by our previous work for three-dimensional free-surface flows \cite{pethiyagoda14a,pethiyagoda14b,pethiyagoda17}, which demonstrated an impressive versatility in terms of computing solutions for a range of disturbances (such as flow past a pressure or a submerged singularity). Many of the ideas used here carry over from our previous work such as the use of a boundary-integral formulation, the implementation of the JFNK method, the general structure of the Jacobian (comprising four submatrices) and the advantage of GPU computing.

We were also motivated by Gardiner et al.~\cite{gardiner15}, who developed a numerical scheme similar to implementation 2 for the problem of computing steady finger solutions in a Hele-Shaw channel. These authors use a conformal mapping approach to form the boundary integral equation by mapping the fluid domain to the infinite strip and then to the half-plane, which is similar to our finite depth examples in Section~\ref{2Dsec:finiteDepth}. However, Gardiner et al.~perform multiple coordinate transforms to the integral to achieve mesh refinement near the boundaries. These transformations slightly change the structure of the Jacobian. Due to the different structure of the Jacobian, Gardiner et al.~chose to keep a few of the rightmost columns along with a chosen bandwidth about the main diagonal when forming a sparse preconditioner. The coordinate transformations performed were specific to the Hele-Shaw problem considered by Gardiner et al.~and thus diminished the generality of their numerical scheme.

Returning to our study, we have considered two-dimensional free-surface flow problems in either finite or infinite depth. All of the problems presented involve conformally mapping the fluid domain to the half-plane, although the finite-depth examples involve an intermediate map to an infinite strip.  In both the infinite-depth and finite-depth cases, the associated Jacobian is primarily comprised of three sparse submatrices and one dense submatrix, and in the case of finite depth flows five additional submatrices with one dimension (row or column) being small. The magnitude of the elements of the dense submatrix are greatest close to the diagonal and decrease away from the diagonal.  For infinite depth flows, the elements decay to zero above and below the diagonal.  For finite depth flows, the elements approach zero above the diagonal and a constant below the diagonal (for $\beta=-1$) or vice versa (for $\beta=1$). The difference between the dense submatrices is not significant enough to require altering our numerical scheme.

It is worth noting that it is possible to approach our two-dimensional free-surface flow problems by formulating an integral equation using Green's second identity (as is done for the three-dimensional analogue~\cite{forbes89,parau02,pethiyagoda14a}) instead of following the standard approach using conformal mapping and complex variable theory.  The relevant integral equation in that case contains a logarithmic term in the integral.  While not shown here, a consequence of this logarithm is that the elements of the dense submatrix of the Jacobian do not decay away from the diagonal, inhibiting the effectiveness of a banded preconditioner.  Thus, in order to implement our approach effectively, the complex variable approach must be used to formulate the boundary integral equation.

\subsection{Benefits of our JFNK scheme}
We have proposed a general banded preconditioner that can be used with the JFNK method to generate solutions with a much larger number of collocation points than presented in the literature.  The larger number of collocation points facilitates more accurate solutions using a scheme that is notorious for not being accurate.  We have computed solutions for flow under a semi-infinite plate or past a pressure distribution in infinite depth (Section \ref{2Dsec:infinitedepth}), along with solutions for flow past a bottom topography in finite depth (Section~\ref{2Dsec:finiteDepth}), to show the adaptability of our numerical scheme to a variety of different two-dimensional free-surface flow problems.

By way of examples, we have demonstrated the advantages of employing a very large number of grid points for solving two-dimensional steady free-surface flow problems.  For example, for highly nonlinear solutions that have a train of very steep waves downstream from a disturbance, the increased number of collocation points has allowed for greater definition around the sharp crests of the waves.  On the other hand, for low Froude number regimes, the waves are sinusoidal but extremely small in amplitude with increasingly shorter wavelengths as the Froude number decreases.  It is simply not possible to determine the dependence of wave properties on Froude number for these flows without using a very large number of grid points, thus ensuring there is significant resolution over each wavelength.  Finally, for supercritical flows there can never be a train of waves downstream, and so there is not normally a need for many thousands of grid points.  However, there are highly nonlinear regimes in which there exists multiple solutions (for a fixed set of parameters) whose free surface profiles are very close to each other.  In these cases, there is some significant advantage in using a large of number of grid points in order to effectively resolve these solutions and the differences between them.

\section*{Acknowledgement}
\noindent SWM acknowledges the support of the Australian Research Council via the Discovery Project DP140100933, while RP acknowledges the support of a Lift-off Fellowship from the Australian Mathematical Society.  The authors are grateful for the computational resources and support provided by the High Performance Computing and Research Support (HPC) group at Queensland University of Technology.  Thanks also goes to the anonymous referees, whose comments have improved the paper.

\appendix

\section{Forming the Jacobian}\label{2Dsec:FormJac}
To construct the Jacobian to be used as a preconditioner matrix (section \ref{2Dsec:NumSch}) we first consider the ordering of unknowns (\ref{2Deq:unkowns}) and the equations in
\begin{equation}
\textbf{F}(\textbf{u})=\left[
\begin{aligned}
\textbf{F}_1\\
\textbf{F}_2
\end{aligned}
\right],\label{2Deq:F}
\end{equation}
where $\textbf{F}_1$ and $\textbf{F}_2$ are the discretised forms of Bernoulli's equation (\ref{2Deq:bernComplex}) and the boundary integral equation (\ref{2Deq:intAltered}), respectively. The elements of $\textbf{F}_1$ and $\textbf{F}_2$ are given by
\begin{align}
\textbf{F}_{1_k} =&\,F_L^2\mathrm{e}^{3\tau_{k+1/2}}\frac{\mathrm{d}\tau_{k+1/2}}{\mathrm{d}\phi}+\sin\theta_{k+1/2}+F_L^2\mathrm{e}^{\tau_{k+1/2}}\left.\frac{\mathrm{d}p}{\mathrm{d}\phi}\right|_{\phi=\phi_{k+1/2}},\label{2Deq:F1}\\
\textbf{F}_{2_k}= &\,\tau_{k+1/2}-\frac{1}{\pi}\sum\limits_{i=1}^{N}\frac{\theta_i-\theta_{k+1/2}}{\phi_i-\phi_{k+1/2}}-\frac{\theta_{k+1/2}}{\pi}\ln\left|\frac{\phi_N-\phi_{k+1/2}}{\phi_1-\phi_{k+1/2}}\right|,\label{2Deq:F2}
\end{align}
for $k=1,\ldots,N-1$.

The derivatives of (\ref{2Deq:F1}) and (\ref{2Deq:F2}) can now be found with respect to the unknowns $\tau_n$ and $\theta_n$ for $n=2,\ldots,N$, giving
\allowdisplaybreaks
\begin{align}
\frac{\partial\textbf{F}_{1_k}}{\partial\tau_n}&=
\begin{cases}
F_L^2\mathrm{e}^{3\tau_{k+1/2}}\left(\frac{3}{2}\frac{\mathrm{d}\tau_{k+1/2}}{\mathrm{d}\phi}+\frac{1}{\Delta\phi}\right)\frac{F_L^2}{2}\mathrm{e}^{\tau_{k+1/2}}\left.\frac{\mathrm{d}p}{\mathrm{d}\phi}\right|_{\phi=\phi_{k+1/2}},
&n=k+1\\
F_L^2\mathrm{e}^{3\tau_{k+1/2}}\left(\frac{3}{2}\frac{\mathrm{d}\tau_{k+1/2}}{\mathrm{d}\phi}-\frac{1}{\Delta\phi}\right)+\frac{F_L^2}{2}\mathrm{e}^{\tau_{k+1/2}}\left.\frac{\mathrm{d}p}{\mathrm{d}\phi}\right|_{\phi=\phi_{k+1/2}},
&n=k\\
0, & \text{otherwise}
\end{cases}\label{2Deq:dF1dTau}\\
\frac{\partial\textbf{F}_{1_k}}{\partial\theta_n}&=
\begin{cases}\displaystyle
\frac{1}{2}\cos\theta_{k+1/2},&n=k,k+1\\
0, & \text{otherwise}
\end{cases}\label{2Deq:dF1dTheta}\\
\frac{\partial\textbf{F}_{2_k}}{\partial\tau_n}&=
\begin{cases}\displaystyle
\frac{1}{2},&n=k-1,k\\
0, & \text{otherwise}
\end{cases}\label{2Deq:dF2dTau}\\
\frac{\partial\textbf{F}_{2_k}}{\partial\theta_n}&=
\begin{cases}
-\frac{1}{\pi}\frac{1}{\phi_n-\phi_{k+1/2}}+\frac{1}{2\pi}\sum\limits_{i=1}^{N}\frac{1}{\phi_i-\phi_{k+1/2}}-\frac{1}{2\pi}\ln\left|\frac{\phi_N-\phi_{k+1/2}}{\phi_1-\phi_{k+1/2}}\right|,
&n=k,k+1\\
\displaystyle-\frac{1}{\pi}\frac{1}{\phi_n-\phi_{k+1/2}}, & \text{otherwise}
\end{cases}\label{2Deq:dF2dTheta}
\end{align}
for $k=1\ldots N-1$ and $n=2\ldots N$. The derivatives (\ref{2Deq:dF1dTau})--(\ref{2Deq:dF2dTheta}) can be arranged in a matrix according to the ordering of unknowns (\ref{2Deq:unkowns}) and equations (\ref{2Deq:F}).

\bibliographystyle{plain}

\end{document}